\documentclass[10pt,english,aps,notitlepage,nofootinbib]{revtex4-1}
\usepackage{amsmath}
\usepackage{esint}
\usepackage{graphicx}
\usepackage{hyperref}
\usepackage{color}
\usepackage[utf8]{inputenc} %%% needed  for some of the references
\newcommand{\be}{\begin{eqnarray}}
\newcommand{\ee}{\end{eqnarray}}
\newcommand{\non}{\nonumber\\}

\newcommand{\ave}[1]{\left\langle #1 \right\rangle}

\newcommand{\gev}{{\rm \, GeV}}
\newcommand{\wn}{N_{\text{part}}}

\begin{document}

\preprint{INT-PUB-16-053; RBRC 1221}
\title{
Correlated stopping, proton clusters and higher order proton cumulants
}

\author{Adam Bzdak}
\email{bzdak@fis.agh.edu.pl}
\affiliation{AGH University of Science and Technology,\\
Faculty of Physics and Applied Computer Science,\\
30-059 Krak\'ow, Poland}

\author{Volker Koch}
\email{vkoch@lbl.gov}
\affiliation{Nuclear Science Division,\\
Lawrence Berkeley National Laboratory,\\
Berkeley, CA, 94720, USA}

\author{Vladimir Skokov}
\email{vskokov@bnl.gov}
\affiliation{RIKEN/BNL, Brookhaven National Laboratory, Upton, NY 11973, USA}

\begin{abstract}
We investigate possible effects of correlations between stopped nucleons on
higher order proton cumulants at low energy heavy-ion collisions. 
We find that fluctuations of the number of wounded nucleons $N_{\mathrm{part}}$ lead
to rather nontrivial 
dependence of the correlations on the centrality; however, 
this effect is too small to explain the 
large and positive four-proton correlations found in the preliminary
data collected by the STAR collaboration at $\sqrt{s}=7.7$ GeV. 
We further demonstrate that, by taking into account additional proton clustering, 
we are able to qualitatively reproduce the preliminary  experimental data. 
We speculate that this clustering 
may originate either from collective/multi-collision stopping which is expected to be effective 
at lower energies or from a possible first-order phase transition, or
from (attractive) final state interactions.  
To test these ideas we propose to measure a mixed multi-particle correlation between stopped
protons and a produced particle (e.g. pion, antiproton). 
\end{abstract}

\maketitle

\section{Introduction}

The structure of the phase diagram is one of the fundamental problems of the
theory of strong interactions, quantum chromodynamics (QCD); a variety of 
solutions to this problems are pursued in both theoretical
and experimental studies. 
From the theory side, the thermodynamics of QCD is
explored by the number of approaches including first principle numerical lattice QCD 
(LQCD)~\cite{Borsanyi:2010cj,Endrodi:2011gv,Bazavov:2011nk,Borsanyi:2012ve,Bellwied:2013cta,Borsanyi:2013bia,Bhattacharya:2014ara,Borsanyi:2014ewa,Bazavov:2014pvz,Ding:2015ona,Bellwied:2015lba,Aarts:2015tyj,Ratti:2016jgx}, 
functional methods~\cite{Fischer:2014mda,Herbst:2013ufa,Mitter:2014wpa,Chatterjee:2015oka} and effective models of QCD~\cite{Fukushima:2003fw,Roder:2003uz,Skokov:2010uh,Friman:2011pf,Pisarski:2016ixt}. 
Experimental studies of the hot and dense
matter created in heavy-ion collisions are also ongoing~\cite{Adamczyk:2013dal} or planned 
to be carried out in the near future~\cite{Chattopadhyay:2016qhg}. 

The LQCD results show that, at the physical pion mass, hot nuclear matter
exhibits residual properties of both dynamical chiral symmetry breaking
and confinement at finite temperature. In the LQCD calculations, it was
demonstrated that the transition between hadrons and quark/gluon
degrees of freedom is of crossover type \cite{Aoki:2006we}. At finite baryon densities, progress of
LQCD calculations is impeded by the notorious sign problem. Several attempts to circumvent
the sign problem were carried out~\cite{Fodor:2001pe,D'Elia:2002gd,Allton:2002zi,Fodor:2004nz,Bonati:2014kpa,deForcrand:2010he}. 
Some of these studies indicate the existence of the expected critical point (CP)
at a finite value of the baryon chemical potential~\cite{Fodor:2001pe,Fodor:2004nz}.

In experiment, the fluctuations of the conserved charges are believed 
to be a promising probe of the QCD critical point. Based on universality arguments, 
it was predicted that the higher order cumulants of baryon/charge number
fluctuations are very sensitive to the correlation length, $\xi$,  and thus 
convey the information about the underlying behaviour of the critical mode~\cite{Stephanov:1999zu,Ejiri:2005wq,Stephanov:2008qz,Stephanov:2011pb}.  
Quantitatively, it was shown that the singular parts of the cumulants of the net baryon/charge 
distribution scale with the correlation length according
to  $\chi^{\rm sing}_n \propto \xi ^ {\frac{n\beta \delta}{\nu}-3}$  
where $\beta$, $\delta$ and $\nu$ are critical exponents of the three-dimensional 
Ising universality class. However, this sensitivity of the higher order cumulants 
to the critical dynamics does not come for free: they probe the 
tails of the probability distribution which
are also susceptible to various non-critical effects  including
baryon number conservation~\cite{Bzdak:2012an}, volume or number of wounded nucleon 
fluctuations~\cite{Skokov:2012ds,Xu:2016qzd,Braun-Munzinger:2016yjz}, detector efficiency and acceptance~\cite{Bzdak:2013pha,Ling:2015yau,Bzdak:2016qdc,Luo:2014rea,Kitazawa:2016awu,Nonaka:2016xje}, 
hadronic rescattering~\cite{Kitazawa:2011wh}, deuteron formation~\cite{Feckova:2015qza}, non-equilibrium effects~\cite{Mukherjee:2015swa,Mukherjee:2016kyu}, 
non-critical correlations between centrality trigger and the observable,  
etc.  

Recent experimental results collected in the  Beam Energy Scan, 
a dedicated experimental program at RHIC,  
demonstrated a non-monotonic dependence of the fourth order cumulant
or kurtosis of net proton fluctuations on the energy of collisions~\cite{Adamczyk:2013dal}. 
Experiments also showed, that, at low energies or, equivalently,  
higher baryon densities, the kurtosis increases with decreasing 
collision energy. In the central rapidity region, where 
most of the measurements are performed, the baryon stopping is 
what makes the high baryon densities possible. Indeed at low energies, 
where proton-antiproton pair-production is negligible, all the observed protons
originate from the target and projectile.
Therefore, the dynamics of 
baryon stopping  is yet another source of fluctuations which can distort the 
measurement and can be potentially responsible for the non-monotonic
behavior of the kurtosis~\cite{Bialas:2016epd}. 

An analysis of the genuine multi-particle correlations \cite{Bzdak:2016sxg} found
that these correlations have a range of the order or larger than  one
unit of rapidity, $\delta y \geq 1$. Typically, long-range correlations in
rapidity suggest the correlations to be formed early in the
evolution of the fireball, although this argument is less stringent at
the lower energies. Therefore, it would be worthwhile to study the effect of
initial conditions, such as so-called volume or rather participant
fluctuations as well as effects of stopping.

It is the purpose of this paper to address some of these issues. We will explore the influence of 
wounded nucleon \cite{Bialas:1976ed} fluctuations as well as fluctuations due to multi-collision 
nucleon stopping, effectively resulting in proton clusters, on the correlation functions at low energies. 
We show that  
for the integrated $n$-particle correlation functions, $C_n$, 
the effect of wounded nucleon fluctuations is small for $n>2$ and negligible for $C_4$. 
However, for $C_2$, we find that 
this contribution is quite significant. 
Additionally,  we show that taking into account multi-proton clusters
reproduces the right order of magnitude of the experimental values for $C_n$. 

This manuscript is organized as follows: 
In Sec.~\ref{Sec:Notation}, we briefly review the definition of the 
correlation functions previously discussed in Ref.~\cite{Bzdak:2016sxg}.
In Sec.~\ref{Sec:WoundN}, we explore the effect of the wounded 
nucleon fluctuations. Additional sources of correlations due to 
proton clustering is discussed in Sec.~\ref{Sec:Multi}. 
We conclude with Sec.~\ref{Sec:Conclusion}.

\section{Notation}
\label{Sec:Notation}
Let us start by defining our notation. 
The  proton multiplicity distribution  measured in a given
rapidity bin, $\Delta y$, will be denoted by $P(N)$.
The corresponding generating function $H(z)$ is given by
\begin{equation}
H(z)=\sum\nolimits_{N}P(N)z^{N},\qquad H(1)=1;
\end{equation}%
so that the factorial moments $F_{k}$ of the proton multiplicity 
distribution can be obtained by taking the appropriate number of derivatives of $H(z)$: 
\begin{equation}
F_{k}\equiv \left\langle N(N-1)\cdots (N-k+1)\right\rangle =\left. \frac{%
d^{k}}{dz^{k}}H(z)\right| _{z=1}.
\end{equation}%
We note that the first factorial moment corresponds to the average
number of particles, $F_{1}=\langle N\rangle$, the second
factorial moment $F_{2}=\langle N(N-1)\rangle$ gives the number of pairs, $%
F_{3}$ the number of triplets etc.
The integrated $k$-particle correlation function $C_{k}$, 
also known as the factorial cumulant
$\kappa_{k}$ \cite{Ling:2015yau}, is given by
\begin{equation}
C_{k}=\left. \frac{d^{k}}{dz^{k}}C(z)\right| _{z=1},\qquad C(z)=\ln \left[
H(z)\right] ,  \label{Cn-def}
\end{equation}%
with  $C_{1}=\left\langle N\right\rangle $. 
As an example, consider  
the two-particle rapidity correlation
function
\begin{eqnarray}
C_{2} &=&\left\langle N(N-1)\right\rangle -\left\langle N\right\rangle ^{2} 
\notag \\
&=&\int_{\Delta y}dy_{1}dy_{2}\left[ \rho _{2}(y_{1,}y_{2})-\rho_{1} (y_{1})\rho_{1}
(y_{2})\right]  \notag \\
&=&\int_{\Delta y}dy_{1}dy_{2}C_{2}(y_{1},y_{2});
\end{eqnarray}%
this can  be indeed recognized as the proper definition of the integrated two-particle 
correlation function.  Here $C_{2}(y_{1},y_{2})$ is the differential
correlation function in rapidity, and $\rho _{2}(y_{1,}y_{2})$ and 
$\rho_{1}(y)$ are the two-particle and the single-particle
rapidity distributions, respectively.

It is convenient to define the reduced correlation functions $c_{k}$
which, following Ref.~\cite{Bzdak:2016sxg}, we shall refer 
to as couplings%
\begin{equation}
c_{k}=\frac{C_{k}}{\left\langle N\right\rangle ^{k}}.  \label{cn-def}
\end{equation}

Finally, the proton cumulants, $K_{n}$, as recently measured by the STAR Collaboration%
\footnote{Note, that the 
STAR collaboration denotes cumulants by $C_{n}$ which we reserve for the correlations.},
are related to the integrated correlation functions $C_{n}$ through%
\begin{eqnarray}
K_{1} &\equiv &\langle N\rangle =C_{1},  \notag \\
K_{2} &\equiv &\langle (\delta N)^{2}\rangle =\left\langle N\right\rangle
+C_{2},  \notag \\
K_{3} &\equiv &\langle \left( \delta N\right) ^{3}\rangle =\left\langle
N\right\rangle +3C_{2}+C_{3},  \notag \\
K_{4} &\equiv &\langle \left( \delta N\right) ^{4}\rangle -3\langle (\delta
N)^{2}\rangle ^{2}=\left\langle N\right\rangle +7C_{2}+6C_{3}+C_{4},
\label{Kn}
\end{eqnarray}%
where $\delta N=N-\langle N\rangle $.\footnote{%
For completeness, let us add that that $K_n$ can be obtained
as well from the above generating function $K_{n}=\left. \frac{d^{n}}{dt^{n}}\ln %
\left[ H(e^{t})\right] \right| _{t=0}.$} 
As was recently emphasized in Refs. \cite{Ling:2015yau,Bzdak:2016sxg}, the cumulants
mix the correlation functions of different orders.

In Ref. \cite{Bzdak:2016sxg},  the correlation functions $C_{n}$ 
and the couplings $c_{n}$ were extracted from the preliminary
STAR data~\cite{Luo:2015ewa,Luo:2015doi}. 
Here we highlight  the most
important conclusions from this analysis.  
For peripheral collisions at $\sqrt{s}=7.7$ GeV, 
the couplings $c_{k}$ scale like $%
1/N^{k-1}$; this is consistent with the production from independent sources.
At $N_{\text{part}}\sim200,$ the correlations $C_{3}$ and $C_{4}$ change signs and reach
large values for the most central collisions   
\begin{equation}
6C_{3}\sim -60, \quad 6c_{3}\sim -10^{-3}\,; \quad\quad C_{4}\sim 170, \quad c_{4}\sim 10^{-4},
\label{eq:c34star}  
\end{equation}
both with large error bars. 
$C_{3}$ and $C_{4}$ decrease with the increasing energy  
and at $\sqrt{s}=19.6$ GeV they are consistent with zero
for $N_{\text{part}}\approx 350$ whereas $C_{2}$ 
does not vary significantly and approximately equals $7 C_{2}\sim -15$ at all energies.
After these preliminaries let us now discuss the effect of volume or
rather participant fluctuations on the correlation functions.

\section{Correlations from fluctuations of the number of wounded nucleons}
\label{Sec:WoundN}
Fluctuations of the  number of wounded or participating nucleons $\wn$, which in the context
of fluctuation studies are often referred to as 
volume fluctuation \cite{Koch:2001cb,Koch:2008ia,Skokov:2012ds},
constitute one of the more obvious sources for correlations and
fluctuations of the final protons. 
Also, at low energies, where baryon pair-production is
negligible, practically all observed protons at mid-rapidity originate from
stopped target and projectile  nucleons. Since pions are abundant even at $7.7 \gev$,
isospin exchange reactions should be very fast and, thus, the observed protons may originate
equally likely from stopped protons and neutrons
\cite{Kitazawa:2011wh,Kitazawa:2012at}. 
This allows us to formulate the following minimal model which
takes into account fluctuations of the number of wounded nucleons, 
baryon number conservation \cite{Bzdak:2012an} and fast  
isospin exchange:
Based on the Glauber model (see, e.g., Ref.~\cite{Alver:2008aq}), 
a certain centrality selection determines 
the distribution of participating nucleons 
$P\left( N_{\text{part}} \right)$. Given the number of participating
nucleons $\wn$  in an
event, the number of protons $N$ observed in $\Delta y$ then follow a binomial distribution 
$B(N;\wn;p)$, where 
$p=\ave{N}/\ave{\wn}$ is the probability for any wounded nucleon to end up in the
rapidity interval $\Delta y$ as a proton. Here, $\ave{N}$ is the observed
mean number of protons in the rapidity interval $\Delta
y$, and $\ave{\wn}$ is the average
number of wounded nucleons for a given centrality selection. 
Obviously this model assumes that each nucleon stops independently
from the other; this finds some support at
higher energies  \cite{Basile:1982we}. 
Thus, the probability $P(N)$ to observe $N$ protons in
$\Delta y$ is given by%
\begin{equation}
P(N)=\sum\nolimits_{N_{\text{part}}}P(N_{\text{part}})\frac{N_{\text{part}}!%
}{N!(N_{\text{part}}-N)!}p^{N}(1-p)^{N_{\text{part}}-N},
\label{eq:PN}
\end{equation}%
and 
\begin{equation}
H(z)=\sum\nolimits_{N_{\text{part}}}P(N_{\text{part}})\left[ 1-p+pz\right]
^{N_{\text{part}}}.  \label{H-binom}
\end{equation}%

Given the above generating function, the correlation functions $C_{k}$
and couplings $c_{k}$ as defined in Sec. \ref{Sec:Notation} can be
computed (see also the Appendix):
\begin{eqnarray}
c_{2} &=&-\frac{1}{\left\langle \wn \right\rangle } +\frac{\langle \left[
\wn-\left\langle \wn \right\rangle \right] ^{2}\rangle }{\left\langle
\wn \right\rangle ^{2}},  \label{c2} \\
c_{3} &=&\frac{2}{\left\langle \wn \right\rangle ^{2}}+\frac{\langle \left[
\wn -\left\langle \wn \right\rangle \right] ^{3}\rangle -3\langle \left[
\wn -\left\langle \wn \right\rangle \right] ^{2}\rangle }{\left\langle
\wn \right\rangle ^{3}},  \label{c3} \\
c_{4} &=&-\frac{6}{\left\langle \wn \right\rangle ^{3}} + \non
&&\frac{\langle \left[
\wn -\left\langle \wn \right\rangle \right] ^{4}\rangle -3\langle \left[
\wn -\left\langle \wn \right\rangle \right] ^{2}\rangle ^{2}+11\langle \left[
\wn -\left\langle \wn \right\rangle \right] ^{2}\rangle -6\langle \left[
\wn -\left\langle \wn \right\rangle \right] ^{3}\rangle }{\left\langle
\wn \right\rangle ^{4}},  \label{c4}
\end{eqnarray}%
where 
\begin{equation}
%\left\langle w^{n}\right\rangle \equiv 
\left\langle N_{\text{part}%
}^{n}\right\rangle =\sum\nolimits_{N_{\text{part}}}P(N_{\text{part}})N_{%
\text{part}}^{n}\text{.}
\end{equation}

In absence of fluctuations in the number of participants,
that is  $P(N_{\text{part}})=\delta_{N_{\text{part}},\ave{N_{{\text{part}}}}}$
in Eq. \eqref{eq:PN}, the couplings, $c_{k}$, reduce to
\begin{equation}
c_{2} \to -\frac{1}{\left\langle N_{\text{part}}\right\rangle },\quad 
c_{3}\to \frac{2}{\left\langle N_{\text{part}}\right\rangle ^{2}},
\quad c_{4} \to -\frac{6%
}{\left\langle N_{\text{part}}\right\rangle ^{3}}.
\label{eq:coupling_no_volume}
\end{equation}%
In this case, 
the couplings, $c_n$, 
as functions of the order, $n$, alternate in sign and 
are suppressed by powers of the mean number of participants,
$\sim 1/\ave{N_{\text{part}}}^{n-1}$. This behavior is  qualitatively
consistent with the analysis of the preliminary STAR data for
peripheral collisions, $N_{\text{part}}<200$ \cite{Bzdak:2016sxg}.  
As expected, the couplings, Eqs.~(\ref{c2}-\ref{c4}), 
do not depend on the binomial probability
$p=\ave{N}/\ave{\wn}$, because  
(binomial) efficiency corrections do not alter 
the reduced correlations functions~(see e.g. Ref.~\cite{Pruneau:2002yf}). 

\subsection{Monte Carlo calculation}

To calculate the contribution of $N_{\text{part}}$ fluctuations to the
multi-particle correlations $C_{k}$ and the couplings $c_{k}$ we need to
define the centrality of the collision. Following the STAR procedure, see e.g. Ref.~\cite{Luo:2015ewa}, we use the
tightest centrality cuts, that is, we calculate $c_{k}$ and $C_{k}$ at a
given number of produced charged particles (except protons) $N_{\text{ch}}$
in $|\eta |<1$ \cite{Luo:2015ewa}.

In our analysis we first calculate $N_{\text{part}}$ using a standard
Glauber model, see, e.g. Ref.~\cite{Alver:2008aq}. We used $\sigma _{\text{in}%
}=31$ mb for $\sqrt{s}= 7.7$ GeV.\footnote{We checked that our
results are insensitive to small variations of $\sigma _{\text{in}}$.}
Next, for each $N_{\text{part}}$ we sampled charged particles from the
Poisson distribution\footnote{%
Usually the number of charged particles is parametrized by the negative
binomial distribution \cite{GrosseOetringhaus:2009kz}. However, we have
checked that possible deviations from Poisson (which are small for lower
energies \cite{GrosseOetringhaus:2009kz}) are not relevant for our results.}
with the average $\langle N_{\text{ch}}\rangle $ in $|\eta |<1$ given by 
\begin{equation}
\left\langle N_{\text{ch}}\right\rangle |_{N_{\text{part}}}=aN_{\text{part}%
}\left( N_{\text{part}}/2\right) ^{0.1}.
\label{eq:ncharge}
\end{equation}%
Here we take  $a=0.75$ for $\sqrt{s}=7.7$ GeV. 
We verified that small variations in the value of $a$ do not change
our conclusions (see Sec.~\ref{Sec:Conclusion} for further discussion). 
Let us comment on $\left( N_{\text{part}}/2\right) ^{0.1}$. It is
well known that in the midrapidity region, the number of charged particles
grows a bit faster than $N_{\text{part}}$, see e.g. Refs.~\cite{Adcox:2000sp,Back:2005hs}.
For example, in Au+Au
collisions at RHIC energies, $\left\langle N_{\text{ch}}\right\rangle /N_{%
\text{part}}$ grows roughly by a factor of $1.7$  
between
proton-proton ($N_{\text{part}}=2$) and central Au+Au collisions ($N_{\text{%
part}}\simeq 350$). 
This feature can be easily understood in the wounded
constituent quark (quark-diquark) model~\cite{Eremin:2003qn,Bialas:2006kw}, where nucleons undergoing several
collisions generate more particles than protons in p+p collisions.

\begin{figure}[t]
\begin{center}
\includegraphics[scale=0.29]{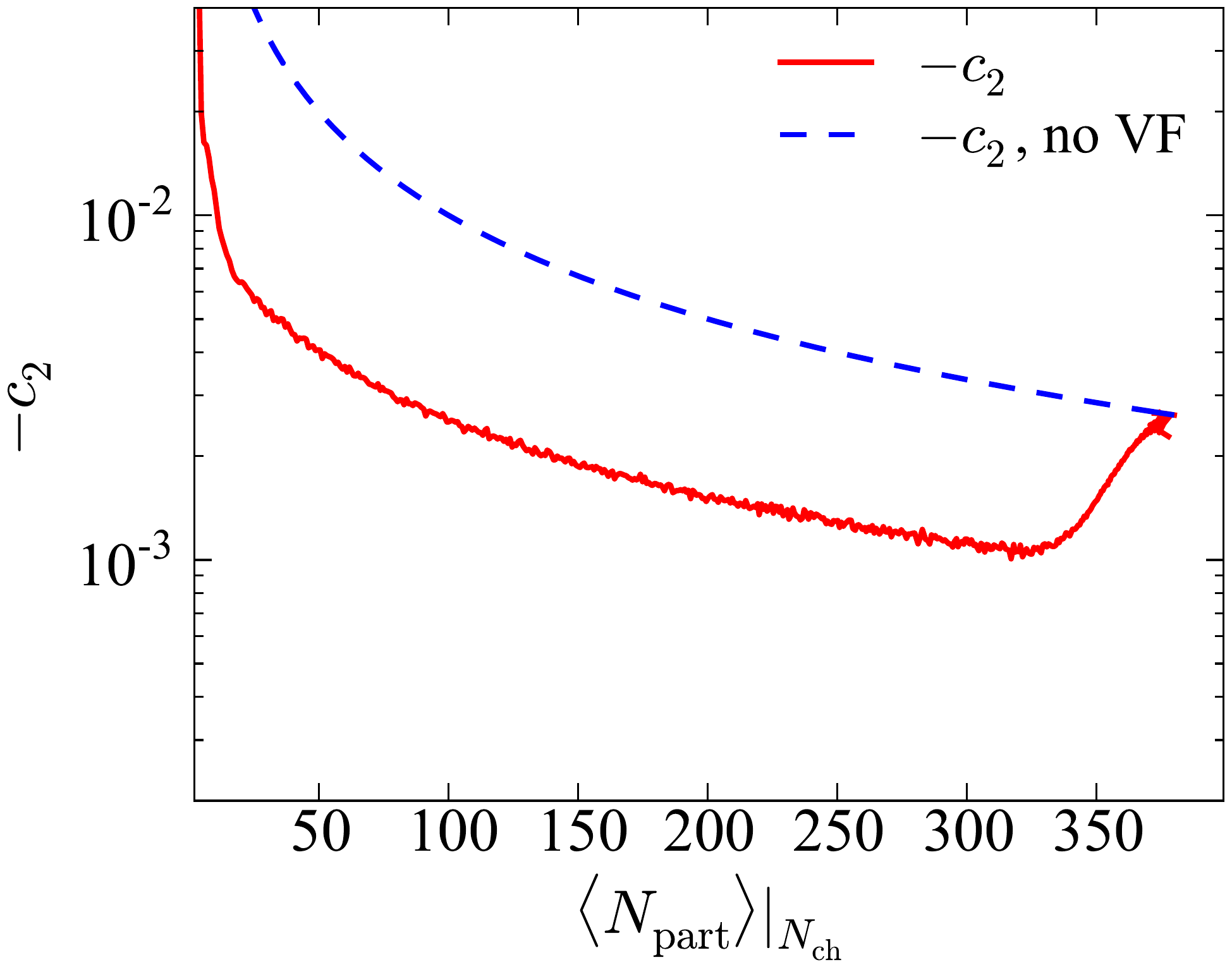} \hspace{0.1cm} %
\includegraphics[scale=0.29]{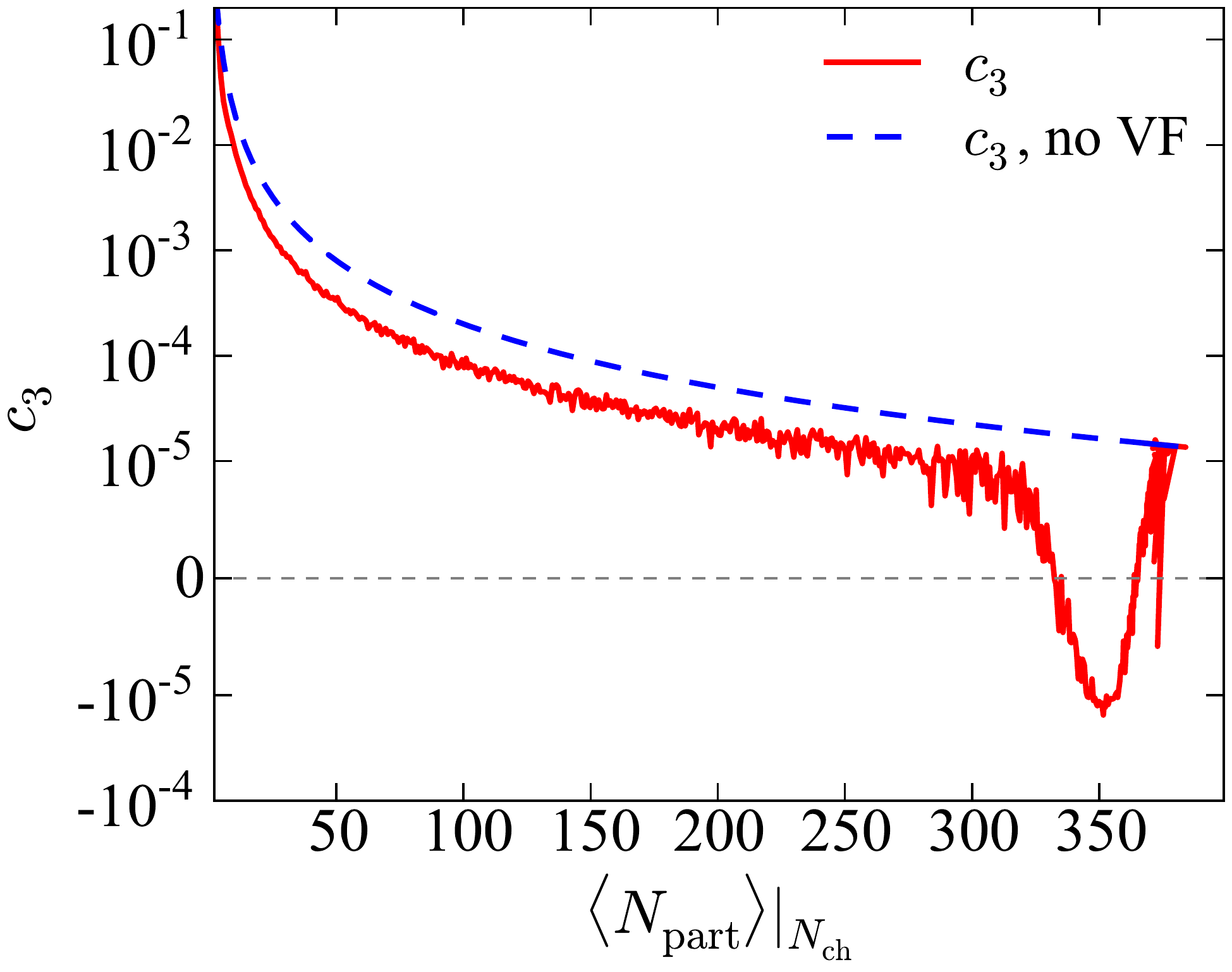} \hspace{0.1cm} %
\includegraphics[scale=0.29]{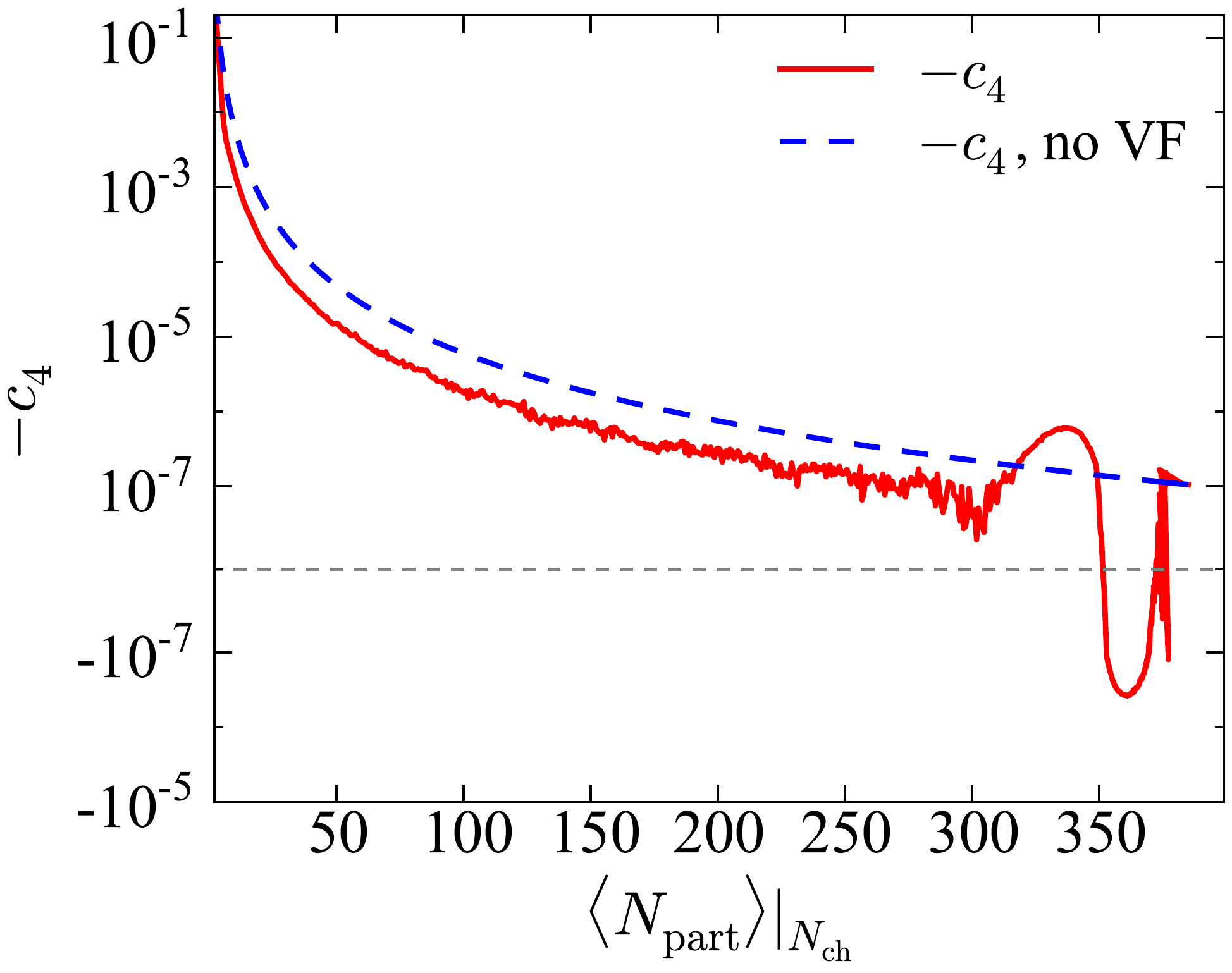}
\end{center}
\par
\vspace{-5mm}
\caption{The couplings $-c_{2},c_{3},-c_{4}$ for $\protect\sqrt{s}=7.7$ GeV in Au+Au
collisions as a function of $\langle N_{\text{part}}\rangle |_{N_{\text{ch}%
}} $ at a given number of charged particles $N_{\text{ch}}$. 
The (blue) dashed curves represent the results {\em without}
fluctuations of the number of wounded nucleons (``no VF''),
see Eq.~\eqref{eq:coupling_no_volume}. 
}
\label{fig:c234-7}
\end{figure}

After generating a sufficient number of events, for each value of $N_{\text{ch}}$
we calculate $\left\langle N_{\text{part}}\right\rangle ,$ $%
\left\langle N_{\text{part}}^{2}\right\rangle $ etc. and evaluate the
couplings $c_{k}$ following Eqs.~(\ref{c2}-\ref{c4}). Our results for $c_{2}$, $c_{3}$,
and $c_{4}$ are shown in Fig.~\ref{fig:c234-7} by the solid (red) curves. 
The (blue) dashed curves represent the results {\em without}
volume ($\wn$) fluctuations (no VF),
see Eq.~\eqref{eq:coupling_no_volume}.  

We observe that $N_{\mathrm{part}}$ fluctuations lead to rather nontrivial
effects in very central collisions.  The coupling $-c_{2}$ changes the trend from decreasing to
increasing with growing $N_{\text{ch}}$ (mind the minus sign in front of $c_{2}$ 
plotted in in Fig.~\ref{fig:c234-7}), 
and both $c_{3}$ and $c_{4}$ change
signs. Interestingly, similar trends were observed for $c_{k}$ extracted
from the preliminary STAR data, see Ref. \cite{Bzdak:2016sxg}, although the
effects for $c_{3}$ and $c_{4}$ observed in Fig. \ref{fig:c234-7} are 
at least an order of magnitude too small for $c_3$, and roughly three orders of magnitude too small for $c_4$, see Eq. (\ref{eq:c34star}). 
Moreover, in our results the
signs change only for very central collisions whereas in the analysis of the
preliminary data this change is present at about $N_{\text{part}}\sim
200$. Finally, as we shall discuss below, the sharp wiggles observed
in $c_{3}$ and $c_{4}$ disappear once one averages the couplings over
a centrality region of $5\%$, as it is done in the STAR
analysis \cite{Luo:2015ewa}.

\begin{figure}[h]
\begin{center}
\includegraphics[scale=0.32]{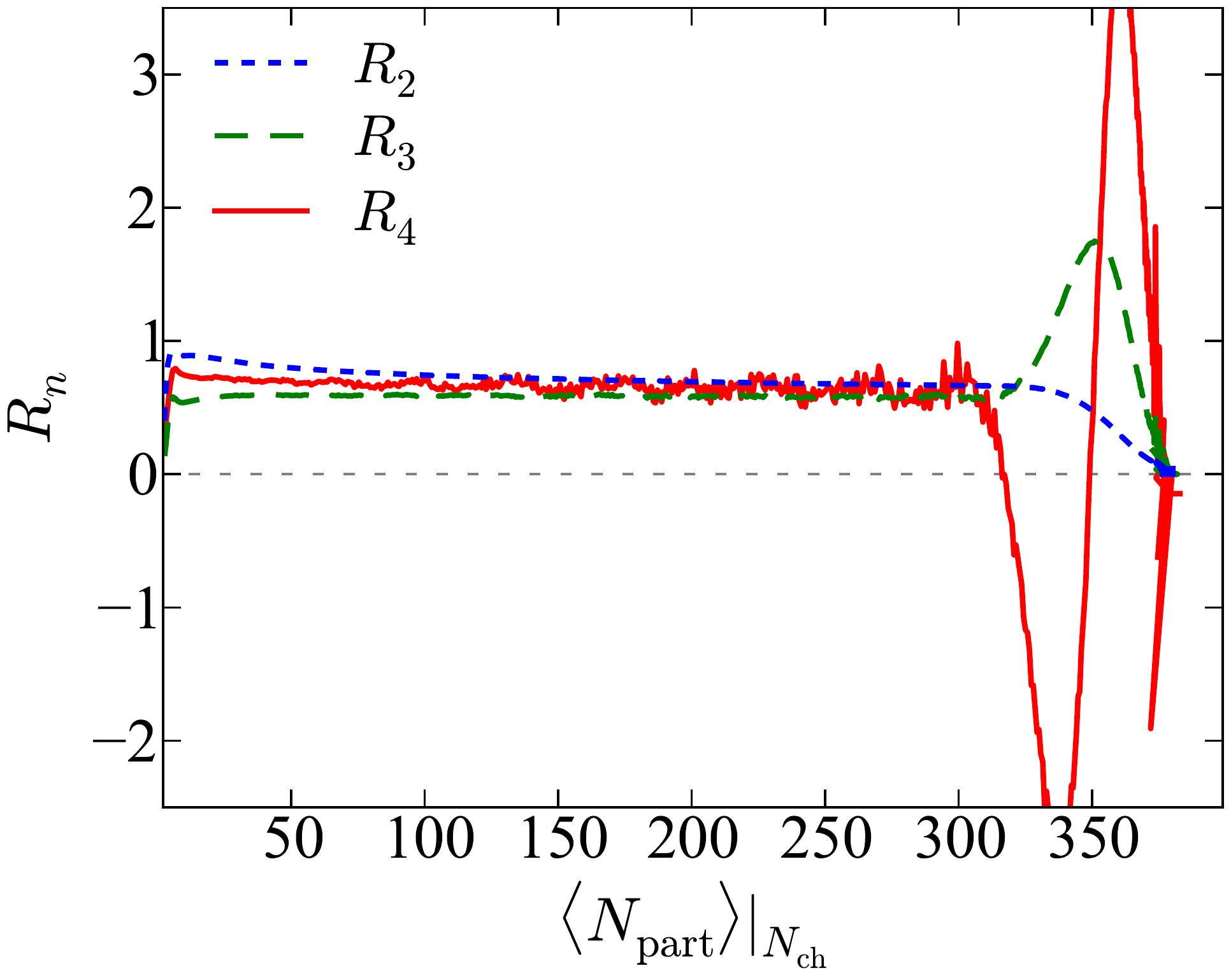} \hspace{0.5cm} %
\end{center}
\par
\vspace{-5mm}
\caption{Relative contribution of $N_{\text{part}}$ fluctuation, see Eqs. (%
\ref{c2R},\ref{c3R},\ref{c4R}), for $c_{2},c_{3}$ and $c_{4}$ at $\protect%
\sqrt{s}=7.7$ GeV in Au+Au collisions. }
\label{fig:correction}
\end{figure}

In order to illustrate the contribution from $N_{\text{part}}$
fluctuations let us factor out the leading term $c_{k} \sim 1/\left\langle
N_{\text{part}}\right\rangle ^{k-1}$ from $c_{2}$, $c_{3}$ and
$c_{4}$ in Eqs.~(\ref{c2}-\ref{c4}), that is define $R_n$ according to 
\begin{eqnarray}
-\frac{1}{\left\langle \wn \right\rangle }\left(
          1-R_{2}\right) &\overset{\mathrm{def}}{=}& c_2,  \label{c2R}\\ 
% ,\quad
% \quad R_{2}=\frac{\langle \left[ \wn -\left\langle \wn \right\rangle \right]
% ^{2}\rangle }{\left\langle \wn \right\rangle },  \label{c2R} \\
\frac{2}{\left\langle \wn\right\rangle ^{2}}\left( 1-R_{3}\right)
&\overset{\mathrm{def}}{=}&
c_{3}, \label{c3R}\\
% ,\quad \quad R_{3}=\frac{3\langle \left[ w-\left\langle w\right\rangle %
% \right] ^{2}\rangle -\langle \left[ w-\left\langle w\right\rangle \right]
% ^{3}\rangle }{2\left\langle w\right\rangle },  \label{c3R} \\
-\frac{6}{\left\langle \wn \right\rangle ^{3}}
          \left(1-R_{4}\right)
	&\overset{\mathrm{def}}{=}& c_{4}	  
		  , \label{c4R}
% ,\quad \quad R_{4}=\frac{\langle \left[ w-\left\langle w\right\rangle \right]
% ^{4}\rangle -3\langle \left[ w-\left\langle w\right\rangle \right]
% ^{2}\rangle ^{2}+11\langle \left[ w-\left\langle w\right\rangle \right]
% ^{2}\rangle -6\langle \left[ w-\left\langle w\right\rangle \right]
% ^{3}\rangle }{6\left\langle w\right\rangle },  \label{c4R}
\end{eqnarray}%
so that $R_{2}$, $R_{3}$ and $R_{4}$ represent the ratios of the
contributions from $\wn$ fluctuations over those arising without
$\wn$ fluctuations in Eq.~\eqref{eq:coupling_no_volume}. In Fig.~\ref{fig:correction} we plot these ratios
as functions of $\left\langle N_{\text{part}}\right\rangle |_{N_{\text{ch}}}$. 
Even though we apply the 
tightest centrality cuts, (we fix the
number of charged particles with the finest possible bin width) 
we  find corrections of $50\%$  or more for
off-central collisions and much larger modification in the most central
collisions.

\begin{figure}[h]
\begin{center}
\includegraphics[scale=0.35]{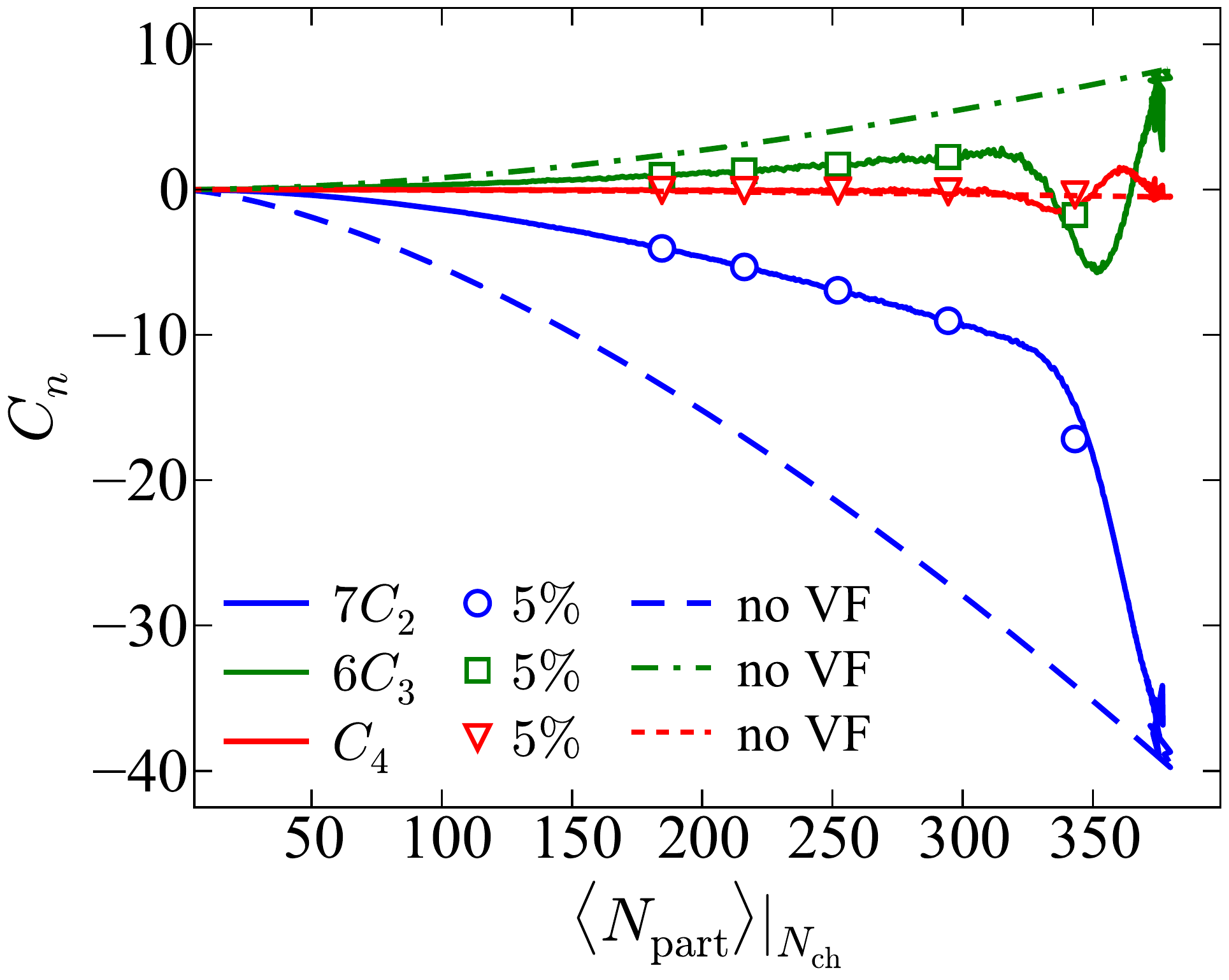}
\hspace{0.5cm} %
\end{center}
\par
\vspace{-5mm}
\caption{Multi-particle correlations $C_{n}$ in Au+Au collisions at $\protect%
\sqrt{s}=7.7$ GeV. 
The leading
terms, where fluctuations of the number of wounded nucleons are not present, are denoted by ``no VF''.
Also shown as circles, triangles and squares are
the results for the five most central bins with a width of $5\%$ of centrality.}
\label{fig:Cn}
\end{figure}

Finally, let us calculate the integrated correlation functions
$C_{k}=\left\langle N\right\rangle ^{k}c_{k}$; they  are directly related to the cumulants measured by STAR,
see Eq.~(\ref{Kn}).
To proceed we need to determine  the dependence of the  average number 
of protons, $\left\langle N\right\rangle $, on $%
N_{\text{part}}$. From the preliminary STAR data
\cite{Luo:2015ewa} we get 
\begin{equation}
\left\langle N\right\rangle |_{N_{\text{ch}}}=b\left( \frac{\left\langle N_{%
\text{part}}\right\rangle |_{N_{\text{ch}}}}{337}\right) ^{1.25},
\end{equation}%
where $b=40$ for $\sqrt{s}=7.7$ GeV. Our conclusions are not sensitive to small variations of $b$ and
changing the exponent from $1.25$ to $1$. 
The results are presented in Fig. \ref{fig:Cn} by the solid curves. The dashed curves correspond
to calculations without volume ($\wn$) fluctuations (no VF). The symbols represent
the correlations after averaging over bins in centrality of
$5\%$, i.e. $0-5\%$, $5-10\%$ etc. Only the five most central points
are shown. For less central collisions, the centrality averaging does not 
alter our results and points fall right on the solid lines.  
Clearly, the contribution originating from $N_{\text{part}}$ fluctuations 
is important for the two particle correlation, $C_{2}$; 
there is also some but less significant effect of $N_{\text{part}}$ fluctuations 
on the three particle correlation $C_{3}$ in central collisions. 
On the other hand, when compared to the STAR data, fluctuations of wounded nucleons  are
all but irrelevant for the four particle correlation, $C_4$. 
In our model calculation,  $C_{4}$  is negative for
off-central collisions and it gets positive for large
$N_{\text{part}}$. After averaging over centrality bins, 
the model predicts around $-0.3$ for $C_{4}$ while the analysis of the
preliminary STAR data gives $\sim 170$. Also, as already mentioned, the
strong oscillations exhibited in $C_{3}$ and $C_{4}$ at large $\wn$
disappear after averaging over centrality bins. 
Obviously our model of independent stopping together with baryon number
conservation clearly fails to explain the preliminary STAR data,
reported in Ref. \cite{Bzdak:2016sxg} (see Fig. 1 therein).

Before we close this section, let us make a few more remarks. First,
the results {\em without} 
the number of wounded nucleon 
fluctuations presented in this section can be verified analytically.
At a fixed $N_{\text{part}}$, Eq. (\ref{H-binom}) reduces to
\begin{equation}
H(z;N_{\text{part}})=(1-p+pz)^{N_{\text{part}}},  \label{H-npart}
\end{equation}%
and using Eq. (\ref{Cn-def}) we obtain%
\begin{equation}
C_{2}=-p^{2}N_{\text{part}},\text{\quad }C_{3}=2p^{3}N_{\text{part}},\quad
C_{4}=-6p^{4}N_{\text{part}}\text{.}  \label{Cn-leading}
\end{equation}
Since $p<1$ this explains the relative magnitude of the
correlation functions. 
Next, in our analysis we assumed that each nucleon is stopped in 
$\Delta y$ with the same probability $p$. This is rather unphysical since
nucleons that collide once are expected to have significantly smaller $p$
than nucleon from the centers which collide several times. However, as
long as we have independent stopping of the nucleons, individual
stopping probabilities do not really change our conclusions. Suppose that each nucleon is
characterized by its own stopping probability, $p_{(i)}$, 
$i=1,...,N_{\text{part}}$. Neglecting $N_{\text{part}}$ fluctuations we obtain at a given $N_{\text{part}}$ 
\footnote{%
The generating function of independent sources is given by a product of its
generating functions.} 
\begin{equation}
H(z;N_{\text{part}})=\prod\nolimits_{i=1}^{N_{\text{part}%
}}(1-p_{(i)}+p_{(i)}z),
\end{equation}%
which obviously reduces to Eq. (\ref{H-npart}) if $p_{i}=p$. Calculating 
$C_{k}$ we observe that it is enough to replace $\wn p^{n}\rightarrow
\sum\nolimits_{i}p_{(i)}^{n}$ in Eq. (\ref{Cn-leading}) and thus the signs
of $C_{k}$ do not change. We conclude that this effect cannot 
lead to a large and positive $C_{4}$ as seen in the STAR data.

The corollary of this section is the following.  
The two-particle correlations obtained in our model of 
independent nucleon stopping together with baryon-number conservation and
fast isospin equilibration 
are of the same magnitude as in the preliminary STAR data. 
Also, the model produces a non-negligible three-particle 
correlation. % this need to be accounted for in a quantitative
             % analysis of the STAR data. 
On the other hand, the model four-particle correlation comes out 
to be almost three orders
of magnitude smaller than in the preliminary STAR
data at $\sqrt{s}=7.7$ GeV.~\footnote{Actually, the presence of this 	
huge discrepancy  is a convincing  demonstration of the
usefulness of the correlation functions. Had we studied the fourth order
cumulant instead, see Eq.~\eqref{Kn}, the model prediction would have
been off only by a factor of $\sim 10$, which of course is still sizable, 
but not as outstanding.}   
The large discrepancy of the four-particle correlation between
our model and the preliminary data suggests that additional strong effects
must be at play. In the following section, we will
explore what happens if we relax the assumption of independent
stopping of nucleons and allow for proton clustering.

\section{Proton clusters}
\label{Sec:Multi}
In this section go we 
go beyond the independent stopping assumption and consider a possible clustering of protons. 
This clustering can be attributed to, e.g., the collective stopping or to 
a first order phase transition. 
We note, that little
is known about the stopping of nucleons at lower energies, and,
therefore, this discussion is somewhat speculative and should be
considered merely as  a motivation to explore the stopping mechanism in more
detail. 

In our view, it is  not at all obvious that the standard Glauber model
still provides the correct stopping framework at energies as low as
$\sqrt{s}=7.7$ GeV. 
Indeed, it seems plausible that a nucleon
passing through a center of a nuclei may actually loose enough energy and
start undergoing elastic scatterings leading to an inter-nucleus cascade.
This may lead to some multi-particle correlations between stopped protons.
For example, in the extreme case when two protons scatter elastically they go back-to-back and
they either both end up in our rapidity bin or they both go
outside. This obviously results in two-particle rapidity correlations
since protons tend to come in pairs. In central collisions more protons can be involved in this
multi-collision (quasi-) elastic stopping, and thus, higher-order proton multiplets
are not a priori excluded. We note that this mechanism is expected to
turn on at relatively low collision energy and in central collisions, where protons
collide several times, and thus may lose sufficient energy to engage
in subsequent elastic scatterings. 
In other words, the
 stopped protons from the centers effectively form proton clusters. Of
 course cluster formation may also arise from a first order
 phase transition (see, e.g., \cite{Skokov:2008zp,Skokov:2009yu,Steinheimer:2012gc}) or final
 state interactions and it will
 require more detailed study to disentangle the various possibilities.   

We  explored various scenarios where protons come in pairs, triplets and
higher-order multiplets. We found that it is not an easy task to
reproduce $C_{2}<0$, $C_{3}<0$ and $C_{4}>0$ with $6C_{3}$ and $C_{4}$ of the
order of $100$ for $\sqrt{s}=7.7$ GeV, as seen in
Ref. \cite{Bzdak:2016sxg}. 
To discuss this in more detail, let us focus on central
$\sqrt{s}=7.7$ GeV collisions, were the signal is strongest. In this
case  $\left\langle N\right\rangle \simeq 40$,  $7 C_{2}\simeq -15 $, $6 C_{3}\simeq -60 $,
and $C_{4}\simeq 170$. 
Before we discuss two scenarios which give results in the right
ballpark, let us put the difficulty of the task in
perspective. Consider a system of clusters  distributed according to  a Poisson
distribution which decay into a fixed number of protons, $m$.  
In this case, the generating function is given by
\begin{equation}
  H_{\text{Poisson}}(z) = \exp\left[ \ave{N_{\rm cl}}\left( z^{m}-1 \right) \right], 
\end{equation}
where $\ave{N_{\rm cl}}$ is the average number of clusters. The resulting integrated correlation 
functions $C_{k}$
are given by $C_{k}=\ave{N_{\rm cl}}m!/(m - k)!$ so that for
four-particle 
clusters ($m=4$) we have $C_{4}= 24 \ave{N_{\rm cl}}$. Thus, in order to get
$C_{4}\simeq 170$ we need on average $6-8$ clusters, about 25 of the 40
observed protons would have to originate from clusters, a rather large fraction
indeed. Of course the same cluster model would also lead to positive two- and
three-particle correlations, contrary to what is seen in the data. 

Let us now turn to discuss two scenarios which lead to a correlation
structure in qualitative agreement with the preliminary STAR data. 
Consider central Au+Au collisions and suppose that nucleons
at the surface (and in general those who do not scatter sufficiently often)
are stopped according to a simple string picture, and, thus, follow
the previously discussed model of independent binomial stopping. Next, suppose that 
nucleons from the centers of target and projectile, which scatter several times,
loose most of the energy and engage in elastic scattering and, thus, fall into $%
\Delta y$ in pairs. In this case the generating function is given by%
\begin{equation}
H(z;N_{\text{part}})=\left( 1-p_{1}+p_{1}z\right) ^{N_{\text{part}%
}-2M}\left( 1-p_{2}+p_{2}z^{2}\right) ^{M},
\end{equation}%
where $M$ is the number of pairs. Let us clarify the above formula. $N_{%
\text{part}}-2M$ nucleons either stop in $\Delta y$ with the probability $%
p_{1}$ or they do not with the probability $1-p_{1}$, and the generating
function for each nucleon is simply given by $(1-p_{1})z^{0}+p_{1}z^{1}$. In
the second term we have $M$ pairs which either fall into $\Delta y$ with
probability $p_{2}$ or they do not with the probability $1-p_{2}$. In this
case we either get $0$ or $2$ protons (from each pair) and the pair
generating function is given by $(1-p_{2})z^{0}+p_{2}z^{2}$. We also expect
that $p_{2}$ is much larger than $p_{1}$ since pairs loose energy due to
scatterings and are likely to fall into the midrapidity bin of interest $\Delta y$ (for
STAR $\Delta y\in \lbrack -0.5,0.5]$).

\begin{figure}[t]
\begin{center}
\includegraphics[scale=0.32]{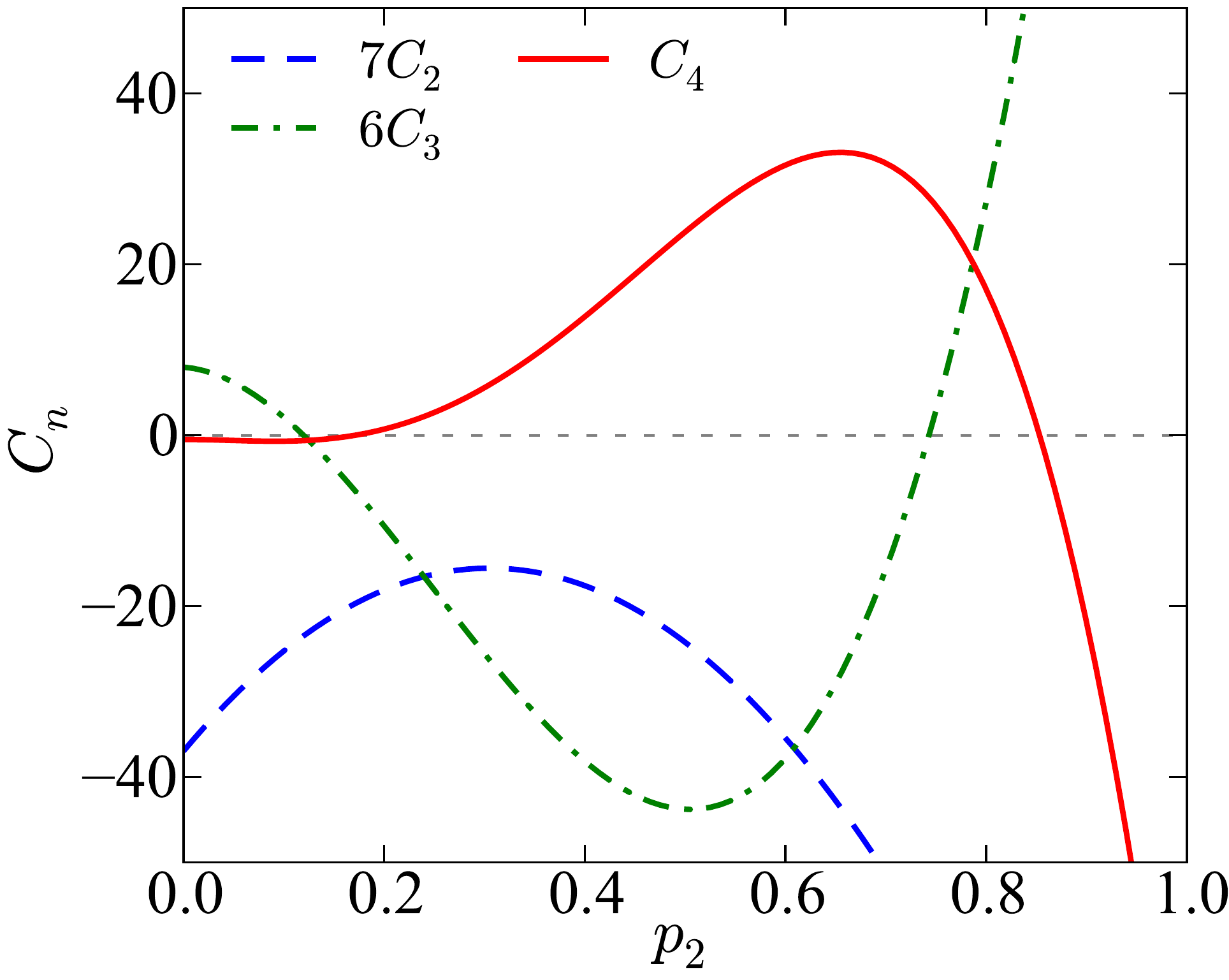} \hspace{0.5cm} %
\includegraphics[scale=0.32]{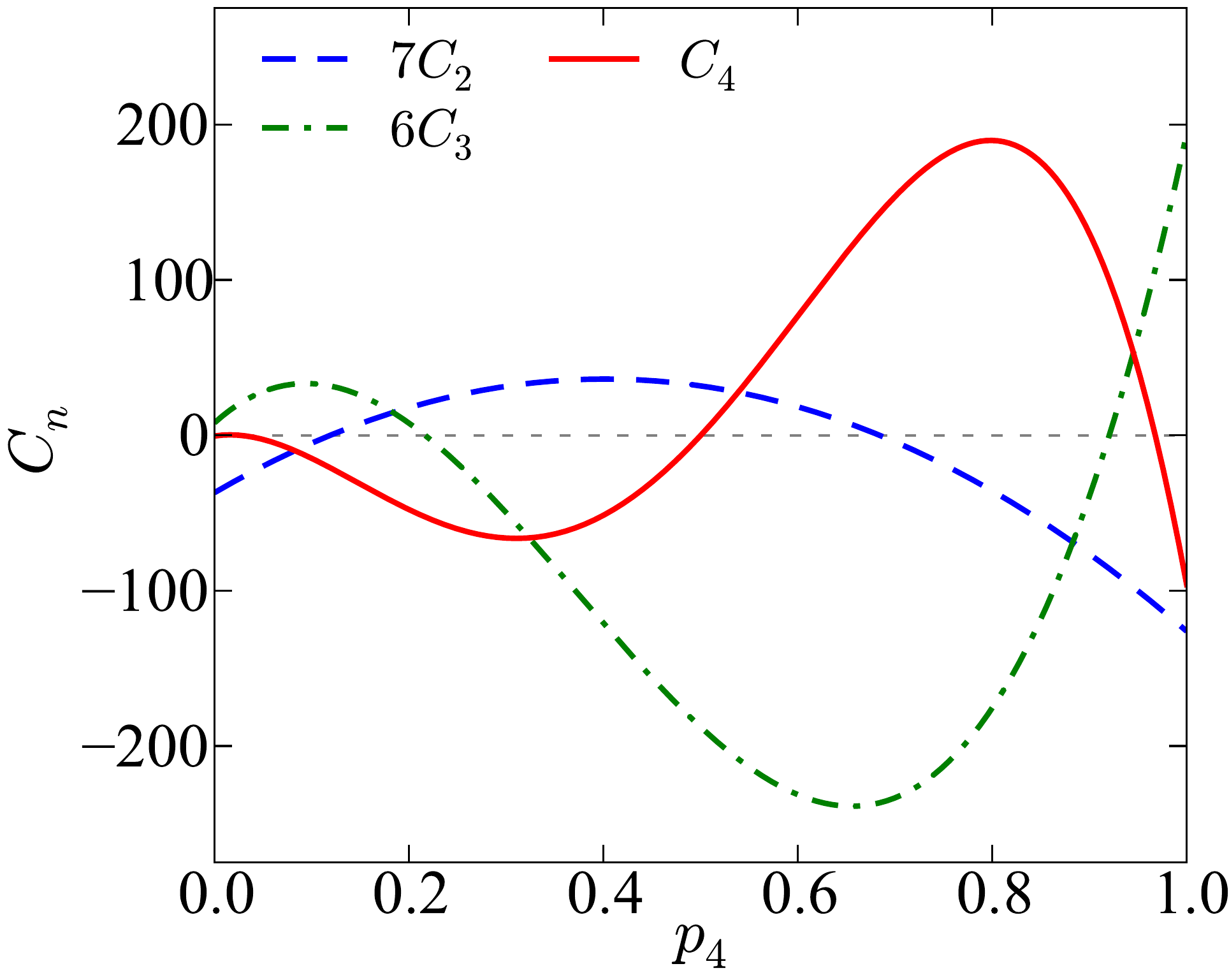}
\end{center}
\par
\vspace{-5mm}
\caption{Integrated multi-particle correlations $C_{n}$ in the model where
particles are correlated in pairs (left) and quartets (right) as a function
of the probability for a pair ($p_{2}$) or a quartet ($p_{4}$) to end up in
the rapidity bin. For larger values of $p_{2}$ and $p_{4}$ we obtain large
values of $C_{3}$ and $C_{4}$. See the text for further explanation.}
\label{fig:pairs}
\end{figure}

The multi-particle correlations $C_{k}$ are given by the appropriate
derivatives (at $z=1$) of 
\begin{eqnarray}
C(z;N_{\text{part}}) &=&\ln \left[ H(z;N_{\text{part}})\right]  \notag \\
&=&(N_{\text{part}}-2M)\ln \left( 1-p_{1}+p_{1}z\right) +M\ln \left(
1-p_{2}+p_{2}z^{2}\right)
\end{eqnarray}%
resulting in 
\begin{eqnarray}
\left\langle N\right\rangle &=&p_{1}\left( N_{\text{part}}-2M\right)
+2p_{2}M,  \notag \\
C_{2} &=&-p_{1}^{2}\left( N_{\text{part}}-2M\right) -2p_{2}\left(
2p_{2}-1\right) M,  \notag \\
C_{3} &=&2p_{1}^{3}\left( N_{\text{part}}-2M\right) -4p_{2}^{2}\left(
3-4p_{2}\right) M,  \notag \\
C_{4} &=&-6p_{1}^{4}\left( N_{\text{part}}-2M\right) +12p_{2}^{2}\left(
8p_{2}-8p_{2}^{2}-1\right) M .
\end{eqnarray}
Taking $N_{\text{part}}=350$, $\left\langle N\right\rangle =0.12N_{\text{part%
}}=42$ and $M=8$ ($8$ pairs of protons) we obtain the relation between $%
p_{1} $ and $p_{2}$. In Fig. \ref{fig:pairs} (left) we plot $7C_{2},$ $6C_{3}
$ and $C_{4}$ as a function of $p_{2}$. We observe that for $p_{2}>0.5$ both 
$C_{3} $ and $C_{4}$ have the right signs and can reach substantial values.

The right panel of Fig.~\ref{fig:pairs} shows the results of an analogues calculation where protons come
in quartets instead of pairs. In this case%
\begin{equation}
C(z;N_{\text{part}})=(N_{\text{part}}-4M)\ln \left( 1-p_{1}+p_{1}z\right)
+M\ln \left( 1-p_{4}+p_{4}z^{4}\right) ,
\end{equation}%
and 
\begin{eqnarray}
\left\langle N\right\rangle &=&p_{1}\left( N_{\text{part}}-4M\right)
+4p_{4}M,  \notag \\
C_{2} &=&-p_{1}^{2}\left( N_{\text{part}}-4M\right) -4p_{4}\left(
4p_{4}-3\right) M,  \notag \\
C_{3} &=&2p_{1}^{3}\left( N_{\text{part}}-4M\right) -8p_{4}\left(
18p_{4}-16p_{4}^{2}-3\right) M,  \notag \\
C_{4} &=&-6p_{1}^{4}\left( N_{\text{part}}-4M\right) +24p_{4}\left(
-34p_{4} + 96p_{4}^{2}-64p_{4}^{3}+1\right) M .
\end{eqnarray}%
where $M$ in this case is the number of proton quartets. In this calculation
we use $M=4$ so that the number of correlated protons is the same as in the
previous case. We observe that the signal for $p_4>0.7$ is much larger, and all $C_n$ agree qualitatively with the STAR data. We have also
verified that the signal increases even further if protons are
clustered in even larger multiplets.

\section{Discussion and conclusions}
\label{Sec:Conclusion}
Let first summarize the main findings of this paper. 
\begin{itemize}
\item We have studied the proton correlations at low energies where proton-antiproton
  pair production can be neglected. To this end we developed a minimal model which is
  based on independent stopping of nucleons, baryon number
  conservation and fast isospin-exchange. We find that this model
  qualitatively reproduces the two-proton correlations seen in the
  preliminary STAR data, while it underpredicts the magnitude of the four-proton 
  correlations by almost three orders of magnitude. 
\item Fluctuations of the  number of participating nucleons, though
  significant even for the tightest centrality cuts, are nowhere near
  large enough to explain the observed four-proton correlations. 
\item The observed large four-particle correlations as well as the
  signs and rough magnitudes of the two- and three-particle correlations
  can be reproduced if one assumes that about 40\% of the observed protons
  originate proton quartets. Given that at lower
  energies the incoming nucleons in the center are likely  to loose
  most of their energy, such a scenario may be not as far fetched as
  one would think initially.  
\end{itemize}

\begin{figure}
\begin{center}
\includegraphics[scale=0.29]{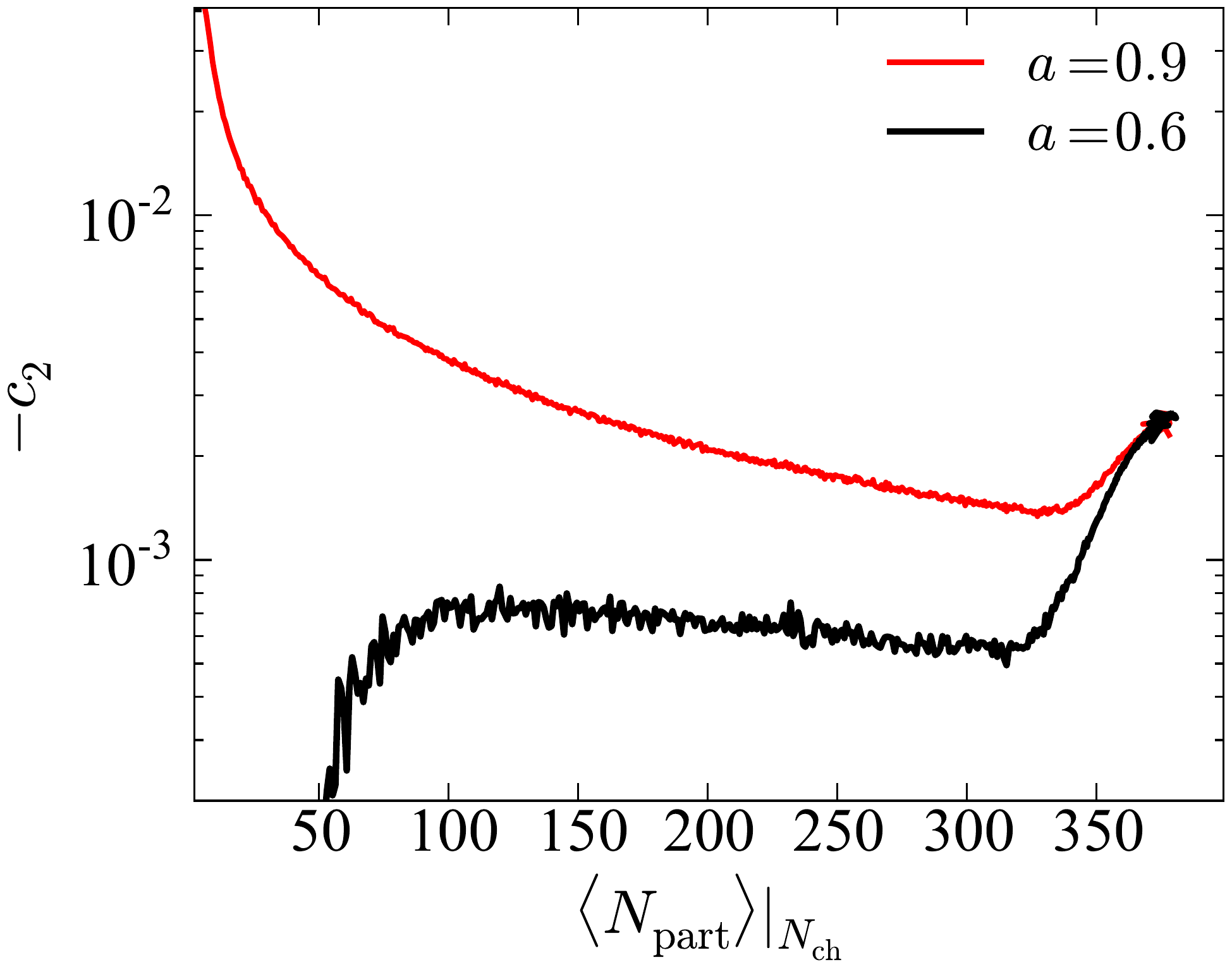} \hspace{0.1cm} %
\includegraphics[scale=0.29]{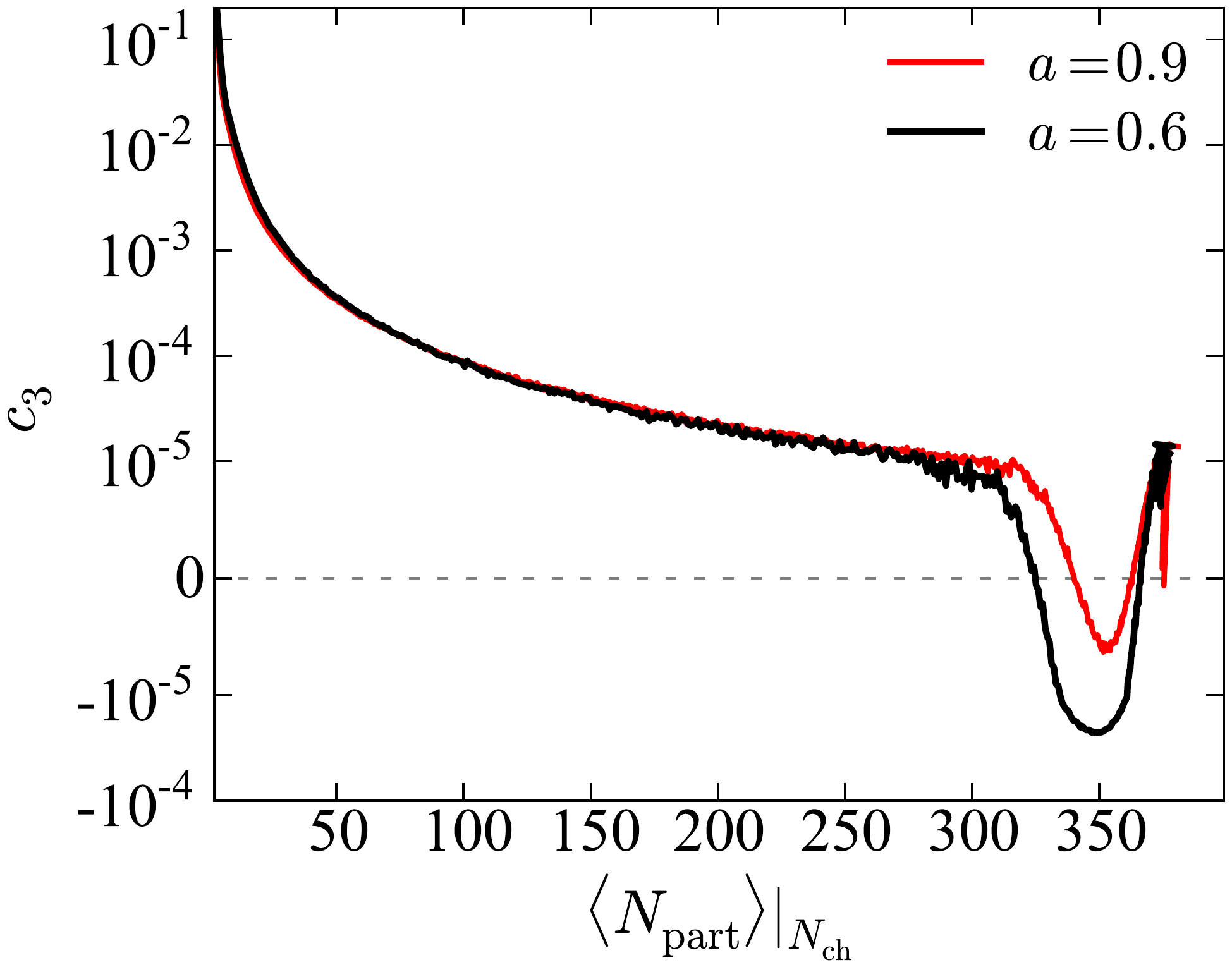} \hspace{0.1cm} %
\includegraphics[scale=0.29]{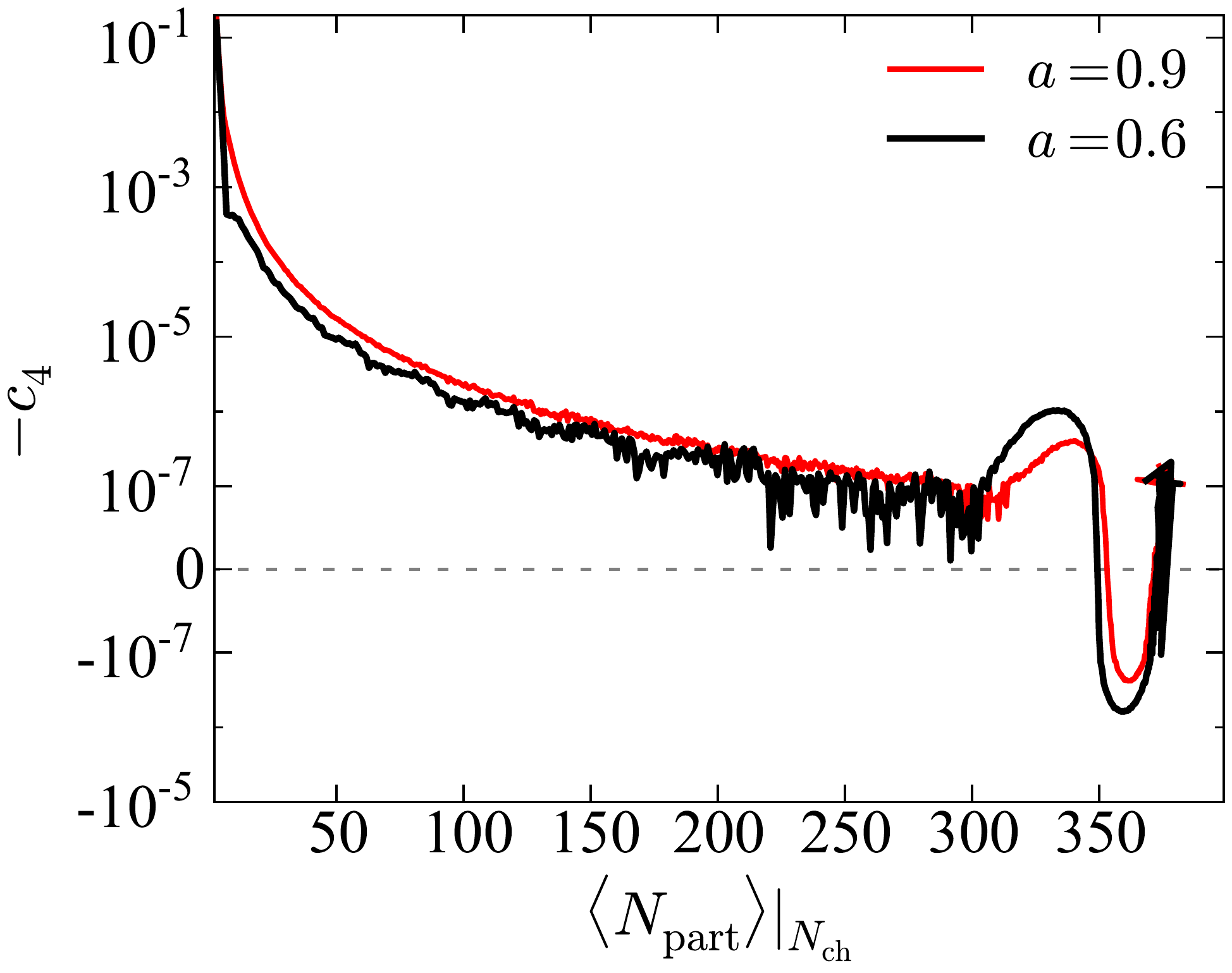}
\end{center}
\par
\vspace{-5mm}
\caption{Same as Fig.~\ref{fig:c234-7} but for different values of
  parameter $a$ of Eq.~\eqref{eq:ncharge}. Instead of $a=0.75$ we use
  $a=0.6$ black lines, and $a=0.9$ red lines.}
\label{fig:c2-dependence}
\end{figure}

Before we conclude, we want to mention that:
\begin{itemize}
\item Although our minimal model fails to reproduce the measured data,
  we consider it as a better baseline for low energies than the  Poisson distribution,
  which is typically used. It respects baryon number conservation, and
  contains the effect of participant fluctuations.   
\item Although we achieved a  qualitative agreement with
  the observed two-particle correlations, the details do matter. 
  This is demonstrated in
  Fig.~\ref{fig:c2-dependence}. There we vary the parameter $a$ of
  Eq. \eqref{eq:ncharge} from its standard value of $a=0.75$ to $a=0.6$ and $a=0.9$ and plot the
  resulting couplings $c_{2}$, $c_{3}$, and $c_4$. While the three- and
  four-particle couplings are hardly changed, the two-particle
  couplings exhibit strong sensitivity on the dependence of the number
  of charged particles on $\wn$. This dependence can only and should  be further 
  constrained by the charged particle distribution, which, however, is not
  yet publicly available for the STAR measurements.
\item 
  Collective stopping or a possible first-order phase transition  may lead 
  to clustering of protons at midrapidity. 
 As we pointed out above, by taking this clustering into account it is possible to qualitatively understand
 the STAR data for the multi-particle correlation functions. 

\item It would be interesting to measure mixed correlations of fourth
  order. For example a correlation $C_{3,1}$ between three protons
  and one produced particle, such as anti-proton or pion may be useful 
  (see the appendix of Ref.~\cite{Bzdak:2016sxg} for the relevant formulas).
  If these
  mixed correlations, or rather couplings $c_{3,1}$, to avoid trivial sensitivity to the number of particles, 
  are as large as the four-proton coupling $c_4$ it
  would rule out collective stopping 
  and provide evidence for a possible first-order phase transition,
  as speculated in Sec.~\ref{Sec:Multi}.
\item
  In this paper we have concentrated on the lowest energy,
  $\sqrt{s}=7.7\gev$. Our basic model equally applies to higher
  energies as long as one can ignore pair production, which is
  probably still the case at $\sqrt{s}=19.6\gev$. There, the observed
  correlations are more or less in agreement with our basic model,
  subject to the aforementioned uncertainty related to the distribution
  of the number of charged particles.  
\end{itemize}

In conclusion, the observed large four-proton correlation exhibited
in the preliminary STAR data at $\sqrt{s}=7.7$ GeV cannot be explained by a combination of
independent baryon stopping, 
participant fluctuations, and baryon-number conservation. However, more
speculative, though not necessarily unrealistic, scenarios involving
the ``collective'' stopping of multiple baryons, may be able to explain
the large part of the observed correlation structure. Therefore, it
is important to test these ideas by, e.g., measuring mixed
correlations. However, given the present analysis and the vary large positive value for the 
four-proton correlation, $C_{4}$, observed in the preliminary STAR data,
it is very difficult to imagine a scenario which explains the data
without employing some kind of cluster formation. If these clusters
originate from collective stopping,  a possible first order phase
transition \cite{Skokov:2008zp,Skokov:2009yu,Steinheimer:2012gc,Steinheimer:2013xxa}, or some
final state effect, remains to be seen, however.  

\appendix

\section{General independent source model}

Here we generalize our Eqs. (\ref{c2},\ref{c3},\ref{c4}) for any independent
sources of protons (see also \cite{Pumplin:1994gh}).

Suppose we have $v$ independent sources of protons distributed according to $%
G(v)$ and each source produces protons according to $p(n_{i})$. The
distribution of measured protons in a given rapidity bin $\Delta y$ is given
by%
\begin{equation}
P(N)=\sum\limits_{v}G(v)\sum_{n_{1},...,n_{v}}p(n_{1})p(n_{2})\cdots
p(n_{v})\delta _{n_{1}+...+n_{v}-N}.
\end{equation}

We note that this formula is quite general and the only assumption we make
is that each source decides on its own how many particles it produces. The
generating function reads%
\begin{eqnarray}
H(z) &=&\sum\nolimits_{N}P(N)z^{N}  \notag \\
&=&\sum\nolimits_{v}G(v)\left[ \sum\nolimits_{n_{1}}p(n_{1})z^{n_{1}}\right]
^{v}  \notag \\
&=&\sum\nolimits_{v}G(v)[H_{1}(z)]^{v},
\end{eqnarray}%
where $H_{1}(z)$ is the generating function for a single source.

To calculate correlation functions we use%
\begin{eqnarray}
C(z) &=&\ln \left[ H(z)\right]  \notag \\
&=&\ln \left[ \sum\nolimits_{v}G(v)e^{v\ln \left[ H_{1}(z)\right] }\right] 
\notag \\
&=&\ln \left[ \sum\nolimits_{v}G(v)e^{vC_{1}(z)}\right] ,
\end{eqnarray}%
where $C_{1}(z)$ characterizes correlations from a single source.

We have%
\begin{equation}
C(1)=C_{1}(1)=0;\quad \frac{dC(z)}{dz}|_{z=1}=\left\langle N\right\rangle
;\quad \frac{dC_{1}(z)}{dz}|_{z=1}=\left\langle n_{1}\right\rangle =\frac{%
\left\langle N\right\rangle }{\left\langle v\right\rangle },
\end{equation}%
where $\left\langle n_{1}\right\rangle $ is the average number of protons
from a single source, $\left\langle v\right\rangle $ is the average numbers
of sources and $\left\langle N\right\rangle $ is the average number of
observed protons.

Calculating derivatives we have for the couplings:%
\begin{eqnarray}
c_{2} &=&\frac{c_{2}^{(1\text{ source})}}{\left\langle v\right\rangle }+%
\frac{\langle \left[ v-\left\langle v\right\rangle \right] ^{2}\rangle }{%
\left\langle v\right\rangle ^{2}},  \label{c2-gen} \\
c_{3} &=&\frac{c_{3}^{(1\text{ source})}}{\left\langle v\right\rangle ^{2}}%
+3c_{2}^{(1\text{ source})}\frac{\langle \left[ v-\left\langle
v\right\rangle \right] ^{2}\rangle }{\left\langle v\right\rangle ^{3}}+\frac{%
\langle \left[ v-\left\langle v\right\rangle \right] ^{3}\rangle }{%
\left\langle v\right\rangle ^{3}},  \label{c3-gen}
\end{eqnarray}%
where $c_{2}^{(1\text{ source})}$ and $c_{3}^{(1\text{ source})}$ are the
couplings for a single source. We note that even even if we have only $2$%
-particle correlations from our sources of protons, they contribute to the
three-particle correlations because of the fluctuating number of sources.

For $c_{4}$ we obtain%
\begin{eqnarray}
c_{4} &=&\frac{c_{4}^{(1\text{ source})}}{\left\langle v\right\rangle ^{3}}+3%
\left[ c_{2}^{(1\text{ source})}\right] ^{2}\frac{\langle \left[
v-\left\langle v\right\rangle \right] ^{2}\rangle }{\left\langle
v\right\rangle ^{4}}+6c_{2}^{(1\text{ source})}\frac{\langle \left[
v-\left\langle v\right\rangle \right] ^{3}\rangle }{\left\langle
v\right\rangle ^{4}}+  \notag \\
&&4c_{3}^{(1\text{ source})}\frac{\langle \left[ v-\left\langle
v\right\rangle \right] ^{2}\rangle }{\left\langle v\right\rangle ^{4}}+\frac{%
\langle \left[ v-\left\langle v\right\rangle \right] ^{4}\rangle -3\langle %
\left[ v-\left\langle v\right\rangle \right] ^{2}\rangle ^{2}}{\left\langle
v\right\rangle ^{4}}.  \label{c4-gen}
\end{eqnarray}

The above formulas are valid for any independent sources of protons. To
obtain Eqs. (\ref{c2},\ref{c3},\ref{c4}) from Sec.~\ref{Sec:WoundN} we note that the
sources of protons are the wounded nucleons themselves. As explained in
Sec.~\ref{Sec:WoundN}, a wounded nucleon appears in our rapidity bin as a proton with
probability $p$ or it does not appear with probability $1-p$. In this case
the multiplicity distribution of protons from a single source is given by%
\begin{equation}
p(n)=\left\{ 
\begin{array}{c}
1-p\text{\quad for\quad }n=0 \\ 
p\quad \text{for\quad }n=1%
\end{array}%
\right. ,
\end{equation}%
and the generating function for a single source is given by%
\begin{equation}
H_{1}(z)=\sum\nolimits_{n}p(n)z^{n}=1-p+pz,
\end{equation}%
which is a known function for binomial distribution. Following Eqs. (\ref%
{Cn-def},\ref{cn-def}) we have%
\begin{equation}
c_{2}^{(1\text{ source})}=-1,\quad c_{3}^{(1\text{ source})}=2,\quad
c_{4}^{(1\text{ source})}=-6.
\end{equation}

Substituting the above values to Eqs. (\ref{c2-gen},\ref{c3-gen},\ref{c4-gen}%
) and putting $v = N_{\text{part}}$, we obtain Eqs. (\ref{c2},\ref{c3},%
\ref{c4}).

 \begin{acknowledgements}
A.B. thanks Larry McLerran for discussions. 
V.S. is indebted to B. Friman and K. Redlich for 
collaborating on the subject of volume fluctuations and their relevance for 
the net baryon cumulants.
We would like to thank the Institute for Nuclear Theory where 
this paper was initiated during the program ``Exploring the QCD Phase Diagram through Energy Scans''. 
A.B. is supported by the Ministry of Science and Higher Education (MNiSW) and 
by the National Science Centre, Grant No. DEC-2014/15/B/ST2/00175, and 
in part by DEC-2013/09/B/ST2/00497.
V.K. was supported by the Office of Nuclear Physics in the US Department of Energy's Office 
of Science under Contract No. DE-AC02-05CH11231.
 \end{acknowledgements}

\bibliography{stop}

\begin{thebibliography}{70}%
\makeatletter
\providecommand \@ifxundefined [1]{%
 \@ifx{#1\undefined}
}%
\providecommand \@ifnum [1]{%
 \ifnum #1\expandafter \@firstoftwo
 \else \expandafter \@secondoftwo
 \fi
}%
\providecommand \@ifx [1]{%
 \ifx #1\expandafter \@firstoftwo
 \else \expandafter \@secondoftwo
 \fi
}%
\providecommand \natexlab [1]{#1}%
\providecommand \enquote  [1]{``#1''}%
\providecommand \bibnamefont  [1]{#1}%
\providecommand \bibfnamefont [1]{#1}%
\providecommand \citenamefont [1]{#1}%
\providecommand \href@noop [0]{\@secondoftwo}%
\providecommand \href [0]{\begingroup \@sanitize@url \@href}%
\providecommand \@href[1]{\@@startlink{#1}\@@href}%
\providecommand \@@href[1]{\endgroup#1\@@endlink}%
\providecommand \@sanitize@url [0]{\catcode `\\12\catcode `\$12\catcode
  `\&12\catcode `\#12\catcode `\^12\catcode `\_12\catcode `\%12\relax}%
\providecommand \@@startlink[1]{}%
\providecommand \@@endlink[0]{}%
\providecommand \url  [0]{\begingroup\@sanitize@url \@url }%
\providecommand \@url [1]{\endgroup\@href {#1}{\urlprefix }}%
\providecommand \urlprefix  [0]{URL }%
\providecommand \Eprint [0]{\href }%
\providecommand \doibase [0]{http://dx.doi.org/}%
\providecommand \selectlanguage [0]{\@gobble}%
\providecommand \bibinfo  [0]{\@secondoftwo}%
\providecommand \bibfield  [0]{\@secondoftwo}%
\providecommand \translation [1]{[#1]}%
\providecommand \BibitemOpen [0]{}%
\providecommand \bibitemStop [0]{}%
\providecommand \bibitemNoStop [0]{.\EOS\space}%
\providecommand \EOS [0]{\spacefactor3000\relax}%
\providecommand \BibitemShut  [1]{\csname bibitem#1\endcsname}%
\let\auto@bib@innerbib\@empty
%</preamble>
\bibitem [{\citenamefont {Borsanyi}\ \emph {et~al.}(2010)\citenamefont
  {Borsanyi}, \citenamefont {Endrodi}, \citenamefont {Fodor}, \citenamefont
  {Jakovac}, \citenamefont {Katz}, \citenamefont {Krieg}, \citenamefont
  {Ratti},\ and\ \citenamefont {Szabo}}]{Borsanyi:2010cj}%
  \BibitemOpen
  \bibfield  {author} {\bibinfo {author} {\bibfnamefont {S.}~\bibnamefont
  {Borsanyi}}, \bibinfo {author} {\bibfnamefont {G.}~\bibnamefont {Endrodi}},
  \bibinfo {author} {\bibfnamefont {Z.}~\bibnamefont {Fodor}}, \bibinfo
  {author} {\bibfnamefont {A.}~\bibnamefont {Jakovac}}, \bibinfo {author}
  {\bibfnamefont {S.~D.}\ \bibnamefont {Katz}}, \bibinfo {author}
  {\bibfnamefont {S.}~\bibnamefont {Krieg}}, \bibinfo {author} {\bibfnamefont
  {C.}~\bibnamefont {Ratti}}, \ and\ \bibinfo {author} {\bibfnamefont {K.~K.}\
  \bibnamefont {Szabo}},\ }\href {\doibase 10.1007/JHEP11(2010)077} {\bibfield
  {journal} {\bibinfo  {journal} {JHEP}\ }\textbf {\bibinfo {volume} {11}},\
  \bibinfo {pages} {077} (\bibinfo {year} {2010})},\ \Eprint
  {http://arxiv.org/abs/1007.2580} {arXiv:1007.2580 [hep-lat]} \BibitemShut
  {NoStop}%
%%CITATION = ARXIV:1007.2580;%%
\bibitem [{\citenamefont {Endrodi}\ \emph {et~al.}(2011)\citenamefont
  {Endrodi}, \citenamefont {Fodor}, \citenamefont {Katz},\ and\ \citenamefont
  {Szabo}}]{Endrodi:2011gv}%
  \BibitemOpen
  \bibfield  {author} {\bibinfo {author} {\bibfnamefont {G.}~\bibnamefont
  {Endrodi}}, \bibinfo {author} {\bibfnamefont {Z.}~\bibnamefont {Fodor}},
  \bibinfo {author} {\bibfnamefont {S.~D.}\ \bibnamefont {Katz}}, \ and\
  \bibinfo {author} {\bibfnamefont {K.~K.}\ \bibnamefont {Szabo}},\ }\href
  {\doibase 10.1007/JHEP04(2011)001} {\bibfield  {journal} {\bibinfo  {journal}
  {JHEP}\ }\textbf {\bibinfo {volume} {04}},\ \bibinfo {pages} {001} (\bibinfo
  {year} {2011})},\ \Eprint {http://arxiv.org/abs/1102.1356} {arXiv:1102.1356
  [hep-lat]} \BibitemShut {NoStop}%
%%CITATION = ARXIV:1102.1356;%%
\bibitem [{\citenamefont {Bazavov}\ \emph {et~al.}(2012)\citenamefont {Bazavov}
  \emph {et~al.}}]{Bazavov:2011nk}%
  \BibitemOpen
  \bibfield  {author} {\bibinfo {author} {\bibfnamefont {A.}~\bibnamefont
  {Bazavov}} \emph {et~al.},\ }\href {\doibase 10.1103/PhysRevD.85.054503}
  {\bibfield  {journal} {\bibinfo  {journal} {Phys. Rev.}\ }\textbf {\bibinfo
  {volume} {D85}},\ \bibinfo {pages} {054503} (\bibinfo {year} {2012})},\
  \Eprint {http://arxiv.org/abs/1111.1710} {arXiv:1111.1710 [hep-lat]}
  \BibitemShut {NoStop}%
%%CITATION = ARXIV:1111.1710;%%
\bibitem [{\citenamefont {Borsanyi}\ \emph {et~al.}(2012)\citenamefont
  {Borsanyi}, \citenamefont {Endrodi}, \citenamefont {Fodor}, \citenamefont
  {Katz},\ and\ \citenamefont {Szabo}}]{Borsanyi:2012ve}%
  \BibitemOpen
  \bibfield  {author} {\bibinfo {author} {\bibfnamefont {S.}~\bibnamefont
  {Borsanyi}}, \bibinfo {author} {\bibfnamefont {G.}~\bibnamefont {Endrodi}},
  \bibinfo {author} {\bibfnamefont {Z.}~\bibnamefont {Fodor}}, \bibinfo
  {author} {\bibfnamefont {S.~D.}\ \bibnamefont {Katz}}, \ and\ \bibinfo
  {author} {\bibfnamefont {K.~K.}\ \bibnamefont {Szabo}},\ }\href {\doibase
  10.1007/JHEP07(2012)056} {\bibfield  {journal} {\bibinfo  {journal} {JHEP}\
  }\textbf {\bibinfo {volume} {07}},\ \bibinfo {pages} {056} (\bibinfo {year}
  {2012})},\ \Eprint {http://arxiv.org/abs/1204.6184} {arXiv:1204.6184
  [hep-lat]} \BibitemShut {NoStop}%
%%CITATION = ARXIV:1204.6184;%%
\bibitem [{\citenamefont {Bellwied}\ \emph {et~al.}(2013)\citenamefont
  {Bellwied}, \citenamefont {Borsanyi}, \citenamefont {Fodor}, \citenamefont
  {Katz},\ and\ \citenamefont {Ratti}}]{Bellwied:2013cta}%
  \BibitemOpen
  \bibfield  {author} {\bibinfo {author} {\bibfnamefont {R.}~\bibnamefont
  {Bellwied}}, \bibinfo {author} {\bibfnamefont {S.}~\bibnamefont {Borsanyi}},
  \bibinfo {author} {\bibfnamefont {Z.}~\bibnamefont {Fodor}}, \bibinfo
  {author} {\bibfnamefont {S.~D.}\ \bibnamefont {Katz}}, \ and\ \bibinfo
  {author} {\bibfnamefont {C.}~\bibnamefont {Ratti}},\ }\href {\doibase
  10.1103/PhysRevLett.111.202302} {\bibfield  {journal} {\bibinfo  {journal}
  {Phys. Rev. Lett.}\ }\textbf {\bibinfo {volume} {111}},\ \bibinfo {pages}
  {202302} (\bibinfo {year} {2013})},\ \Eprint {http://arxiv.org/abs/1305.6297}
  {arXiv:1305.6297 [hep-lat]} \BibitemShut {NoStop}%
%%CITATION = ARXIV:1305.6297;%%
\bibitem [{\citenamefont {Borsanyi}\ \emph
  {et~al.}(2014{\natexlab{a}})\citenamefont {Borsanyi}, \citenamefont {Fodor},
  \citenamefont {Hoelbling}, \citenamefont {Katz}, \citenamefont {Krieg},\ and\
  \citenamefont {Szabo}}]{Borsanyi:2013bia}%
  \BibitemOpen
  \bibfield  {author} {\bibinfo {author} {\bibfnamefont {S.}~\bibnamefont
  {Borsanyi}}, \bibinfo {author} {\bibfnamefont {Z.}~\bibnamefont {Fodor}},
  \bibinfo {author} {\bibfnamefont {C.}~\bibnamefont {Hoelbling}}, \bibinfo
  {author} {\bibfnamefont {S.~D.}\ \bibnamefont {Katz}}, \bibinfo {author}
  {\bibfnamefont {S.}~\bibnamefont {Krieg}}, \ and\ \bibinfo {author}
  {\bibfnamefont {K.~K.}\ \bibnamefont {Szabo}},\ }\href {\doibase
  10.1016/j.physletb.2014.01.007} {\bibfield  {journal} {\bibinfo  {journal}
  {Phys. Lett.}\ }\textbf {\bibinfo {volume} {B730}},\ \bibinfo {pages} {99}
  (\bibinfo {year} {2014}{\natexlab{a}})},\ \Eprint
  {http://arxiv.org/abs/1309.5258} {arXiv:1309.5258 [hep-lat]} \BibitemShut
  {NoStop}%
%%CITATION = ARXIV:1309.5258;%%
\bibitem [{\citenamefont {Bhattacharya}\ \emph {et~al.}(2014)\citenamefont
  {Bhattacharya} \emph {et~al.}}]{Bhattacharya:2014ara}%
  \BibitemOpen
  \bibfield  {author} {\bibinfo {author} {\bibfnamefont {T.}~\bibnamefont
  {Bhattacharya}} \emph {et~al.},\ }\href {\doibase
  10.1103/PhysRevLett.113.082001} {\bibfield  {journal} {\bibinfo  {journal}
  {Phys. Rev. Lett.}\ }\textbf {\bibinfo {volume} {113}},\ \bibinfo {pages}
  {082001} (\bibinfo {year} {2014})},\ \Eprint {http://arxiv.org/abs/1402.5175}
  {arXiv:1402.5175 [hep-lat]} \BibitemShut {NoStop}%
%%CITATION = ARXIV:1402.5175;%%
\bibitem [{\citenamefont {Borsanyi}\ \emph
  {et~al.}(2014{\natexlab{b}})\citenamefont {Borsanyi}, \citenamefont {Fodor},
  \citenamefont {Katz}, \citenamefont {Krieg}, \citenamefont {Ratti},\ and\
  \citenamefont {Szabo}}]{Borsanyi:2014ewa}%
  \BibitemOpen
  \bibfield  {author} {\bibinfo {author} {\bibfnamefont {S.}~\bibnamefont
  {Borsanyi}}, \bibinfo {author} {\bibfnamefont {Z.}~\bibnamefont {Fodor}},
  \bibinfo {author} {\bibfnamefont {S.~D.}\ \bibnamefont {Katz}}, \bibinfo
  {author} {\bibfnamefont {S.}~\bibnamefont {Krieg}}, \bibinfo {author}
  {\bibfnamefont {C.}~\bibnamefont {Ratti}}, \ and\ \bibinfo {author}
  {\bibfnamefont {K.~K.}\ \bibnamefont {Szabo}},\ }\href {\doibase
  10.1103/PhysRevLett.113.052301} {\bibfield  {journal} {\bibinfo  {journal}
  {Phys. Rev. Lett.}\ }\textbf {\bibinfo {volume} {113}},\ \bibinfo {pages}
  {052301} (\bibinfo {year} {2014}{\natexlab{b}})},\ \Eprint
  {http://arxiv.org/abs/1403.4576} {arXiv:1403.4576 [hep-lat]} \BibitemShut
  {NoStop}%
%%CITATION = ARXIV:1403.4576;%%
\bibitem [{\citenamefont {Bazavov}\ \emph {et~al.}(2014)\citenamefont {Bazavov}
  \emph {et~al.}}]{Bazavov:2014pvz}%
  \BibitemOpen
  \bibfield  {author} {\bibinfo {author} {\bibfnamefont {A.}~\bibnamefont
  {Bazavov}} \emph {et~al.} (\bibinfo {collaboration} {HotQCD}),\ }\href
  {\doibase 10.1103/PhysRevD.90.094503} {\bibfield  {journal} {\bibinfo
  {journal} {Phys. Rev.}\ }\textbf {\bibinfo {volume} {D90}},\ \bibinfo {pages}
  {094503} (\bibinfo {year} {2014})},\ \Eprint {http://arxiv.org/abs/1407.6387}
  {arXiv:1407.6387 [hep-lat]} \BibitemShut {NoStop}%
%%CITATION = ARXIV:1407.6387;%%
\bibitem [{\citenamefont {Ding}\ \emph {et~al.}(2015)\citenamefont {Ding},
  \citenamefont {Karsch},\ and\ \citenamefont {Mukherjee}}]{Ding:2015ona}%
  \BibitemOpen
  \bibfield  {author} {\bibinfo {author} {\bibfnamefont {H.-T.}\ \bibnamefont
  {Ding}}, \bibinfo {author} {\bibfnamefont {F.}~\bibnamefont {Karsch}}, \ and\
  \bibinfo {author} {\bibfnamefont {S.}~\bibnamefont {Mukherjee}},\ }\href
  {\doibase 10.1142/S0218301315300076} {\bibfield  {journal} {\bibinfo
  {journal} {Int. J. Mod. Phys.}\ }\textbf {\bibinfo {volume} {E24}},\ \bibinfo
  {pages} {1530007} (\bibinfo {year} {2015})},\ \Eprint
  {http://arxiv.org/abs/1504.05274} {arXiv:1504.05274 [hep-lat]} \BibitemShut
  {NoStop}%
%%CITATION = ARXIV:1504.05274;%%
\bibitem [{\citenamefont {Bellwied}\ \emph {et~al.}(2015)\citenamefont
  {Bellwied}, \citenamefont {Borsanyi}, \citenamefont {Fodor}, \citenamefont
  {Katz}, \citenamefont {Pasztor}, \citenamefont {Ratti},\ and\ \citenamefont
  {Szabo}}]{Bellwied:2015lba}%
  \BibitemOpen
  \bibfield  {author} {\bibinfo {author} {\bibfnamefont {R.}~\bibnamefont
  {Bellwied}}, \bibinfo {author} {\bibfnamefont {S.}~\bibnamefont {Borsanyi}},
  \bibinfo {author} {\bibfnamefont {Z.}~\bibnamefont {Fodor}}, \bibinfo
  {author} {\bibfnamefont {S.~D.}\ \bibnamefont {Katz}}, \bibinfo {author}
  {\bibfnamefont {A.}~\bibnamefont {Pasztor}}, \bibinfo {author} {\bibfnamefont
  {C.}~\bibnamefont {Ratti}}, \ and\ \bibinfo {author} {\bibfnamefont {K.~K.}\
  \bibnamefont {Szabo}},\ }\href {\doibase 10.1103/PhysRevD.92.114505}
  {\bibfield  {journal} {\bibinfo  {journal} {Phys. Rev.}\ }\textbf {\bibinfo
  {volume} {D92}},\ \bibinfo {pages} {114505} (\bibinfo {year} {2015})},\
  \Eprint {http://arxiv.org/abs/1507.04627} {arXiv:1507.04627 [hep-lat]}
  \BibitemShut {NoStop}%
%%CITATION = ARXIV:1507.04627;%%
\bibitem [{\citenamefont {Aarts}(2016)}]{Aarts:2015tyj}%
  \BibitemOpen
  \bibfield  {author} {\bibinfo {author} {\bibfnamefont {G.}~\bibnamefont
  {Aarts}},\ }\bibfield  {booktitle} {\emph {\bibinfo {booktitle}
  {{Proceedings, 13th International Workshop on Hadron Physics: Angra dos Reis,
  Rio de Janeiro, Brazil, March 22-27, 2015}}},\ }\href {\doibase
  10.1088/1742-6596/706/2/022004} {\bibfield  {journal} {\bibinfo  {journal}
  {J. Phys. Conf. Ser.}\ }\textbf {\bibinfo {volume} {706}},\ \bibinfo {pages}
  {022004} (\bibinfo {year} {2016})},\ \Eprint
  {http://arxiv.org/abs/1512.05145} {arXiv:1512.05145 [hep-lat]} \BibitemShut
  {NoStop}%
%%CITATION = ARXIV:1512.05145;%%
\bibitem [{\citenamefont {Ratti}(2016)}]{Ratti:2016jgx}%
  \BibitemOpen
  \bibfield  {author} {\bibinfo {author} {\bibfnamefont {C.}~\bibnamefont
  {Ratti}},\ }\bibfield  {booktitle} {\emph {\bibinfo {booktitle}
  {{Proceedings, 25th International Conference on Ultra-Relativistic
  Nucleus-Nucleus Collisions (Quark Matter 2015): Kobe, Japan, September
  27-October 3, 2015}}},\ }\href {\doibase 10.1016/j.nuclphysa.2016.02.022}
  {\bibfield  {journal} {\bibinfo  {journal} {Nucl. Phys.}\ }\textbf {\bibinfo
  {volume} {A956}},\ \bibinfo {pages} {51} (\bibinfo {year} {2016})},\ \Eprint
  {http://arxiv.org/abs/1601.02367} {arXiv:1601.02367 [hep-lat]} \BibitemShut
  {NoStop}%
%%CITATION = ARXIV:1601.02367;%%
\bibitem [{\citenamefont {Fischer}\ \emph {et~al.}(2014)\citenamefont
  {Fischer}, \citenamefont {Luecker},\ and\ \citenamefont
  {Welzbacher}}]{Fischer:2014mda}%
  \BibitemOpen
  \bibfield  {author} {\bibinfo {author} {\bibfnamefont {C.~S.}\ \bibnamefont
  {Fischer}}, \bibinfo {author} {\bibfnamefont {J.}~\bibnamefont {Luecker}}, \
  and\ \bibinfo {author} {\bibfnamefont {C.~A.}\ \bibnamefont {Welzbacher}},\
  }\bibfield  {booktitle} {\emph {\bibinfo {booktitle} {{Proceedings, 24th
  International Conference on Ultra-Relativistic Nucleus-Nucleus Collisions
  (Quark Matter 2014): Darmstadt, Germany, May 19-24, 2014}}},\ }\href
  {\doibase 10.1016/j.nuclphysa.2014.09.033} {\bibfield  {journal} {\bibinfo
  {journal} {Nucl. Phys.}\ }\textbf {\bibinfo {volume} {A931}},\ \bibinfo
  {pages} {774} (\bibinfo {year} {2014})},\ \Eprint
  {http://arxiv.org/abs/1410.0124} {arXiv:1410.0124 [hep-ph]} \BibitemShut
  {NoStop}%
%%CITATION = ARXIV:1410.0124;%%
\bibitem [{\citenamefont {Herbst}\ \emph {et~al.}(2014)\citenamefont {Herbst},
  \citenamefont {Mitter}, \citenamefont {Pawlowski}, \citenamefont {Schaefer},\
  and\ \citenamefont {Stiele}}]{Herbst:2013ufa}%
  \BibitemOpen
  \bibfield  {author} {\bibinfo {author} {\bibfnamefont {T.~K.}\ \bibnamefont
  {Herbst}}, \bibinfo {author} {\bibfnamefont {M.}~\bibnamefont {Mitter}},
  \bibinfo {author} {\bibfnamefont {J.~M.}\ \bibnamefont {Pawlowski}}, \bibinfo
  {author} {\bibfnamefont {B.-J.}\ \bibnamefont {Schaefer}}, \ and\ \bibinfo
  {author} {\bibfnamefont {R.}~\bibnamefont {Stiele}},\ }\href {\doibase
  10.1016/j.physletb.2014.02.045} {\bibfield  {journal} {\bibinfo  {journal}
  {Phys. Lett.}\ }\textbf {\bibinfo {volume} {B731}},\ \bibinfo {pages} {248}
  (\bibinfo {year} {2014})},\ \Eprint {http://arxiv.org/abs/1308.3621}
  {arXiv:1308.3621 [hep-ph]} \BibitemShut {NoStop}%
%%CITATION = ARXIV:1308.3621;%%
\bibitem [{\citenamefont {Mitter}\ \emph {et~al.}(2015)\citenamefont {Mitter},
  \citenamefont {Pawlowski},\ and\ \citenamefont
  {Strodthoff}}]{Mitter:2014wpa}%
  \BibitemOpen
  \bibfield  {author} {\bibinfo {author} {\bibfnamefont {M.}~\bibnamefont
  {Mitter}}, \bibinfo {author} {\bibfnamefont {J.~M.}\ \bibnamefont
  {Pawlowski}}, \ and\ \bibinfo {author} {\bibfnamefont {N.}~\bibnamefont
  {Strodthoff}},\ }\href {\doibase 10.1103/PhysRevD.91.054035} {\bibfield
  {journal} {\bibinfo  {journal} {Phys. Rev.}\ }\textbf {\bibinfo {volume}
  {D91}},\ \bibinfo {pages} {054035} (\bibinfo {year} {2015})},\ \Eprint
  {http://arxiv.org/abs/1411.7978} {arXiv:1411.7978 [hep-ph]} \BibitemShut
  {NoStop}%
%%CITATION = ARXIV:1411.7978;%%
\bibitem [{\citenamefont {Chatterjee}\ and\ \citenamefont
  {Mohan}(2015)}]{Chatterjee:2015oka}%
  \BibitemOpen
  \bibfield  {author} {\bibinfo {author} {\bibfnamefont {S.}~\bibnamefont
  {Chatterjee}}\ and\ \bibinfo {author} {\bibfnamefont {K.~A.}\ \bibnamefont
  {Mohan}},\ }\href@noop {} {\  (\bibinfo {year} {2015})},\ \Eprint
  {http://arxiv.org/abs/1502.00648} {arXiv:1502.00648 [nucl-th]} \BibitemShut
  {NoStop}%
%%CITATION = ARXIV:1502.00648;%%
\bibitem [{\citenamefont {Fukushima}(2004)}]{Fukushima:2003fw}%
  \BibitemOpen
  \bibfield  {author} {\bibinfo {author} {\bibfnamefont {K.}~\bibnamefont
  {Fukushima}},\ }\href {\doibase 10.1016/j.physletb.2004.04.027} {\bibfield
  {journal} {\bibinfo  {journal} {Phys. Lett.}\ }\textbf {\bibinfo {volume}
  {B591}},\ \bibinfo {pages} {277} (\bibinfo {year} {2004})},\ \Eprint
  {http://arxiv.org/abs/hep-ph/0310121} {arXiv:hep-ph/0310121 [hep-ph]}
  \BibitemShut {NoStop}%
%%CITATION = HEP-PH/0310121;%%
\bibitem [{\citenamefont {Roder}\ \emph {et~al.}(2003)\citenamefont {Roder},
  \citenamefont {Ruppert},\ and\ \citenamefont {Rischke}}]{Roder:2003uz}%
  \BibitemOpen
  \bibfield  {author} {\bibinfo {author} {\bibfnamefont {D.}~\bibnamefont
  {Roder}}, \bibinfo {author} {\bibfnamefont {J.}~\bibnamefont {Ruppert}}, \
  and\ \bibinfo {author} {\bibfnamefont {D.~H.}\ \bibnamefont {Rischke}},\
  }\href {\doibase 10.1103/PhysRevD.68.016003} {\bibfield  {journal} {\bibinfo
  {journal} {Phys. Rev.}\ }\textbf {\bibinfo {volume} {D68}},\ \bibinfo {pages}
  {016003} (\bibinfo {year} {2003})},\ \Eprint
  {http://arxiv.org/abs/nucl-th/0301085} {arXiv:nucl-th/0301085 [nucl-th]}
  \BibitemShut {NoStop}%
%%CITATION = NUCL-TH/0301085;%%
\bibitem [{\citenamefont {Skokov}\ \emph {et~al.}(2011)\citenamefont {Skokov},
  \citenamefont {Friman},\ and\ \citenamefont {Redlich}}]{Skokov:2010uh}%
  \BibitemOpen
  \bibfield  {author} {\bibinfo {author} {\bibfnamefont {V.}~\bibnamefont
  {Skokov}}, \bibinfo {author} {\bibfnamefont {B.}~\bibnamefont {Friman}}, \
  and\ \bibinfo {author} {\bibfnamefont {K.}~\bibnamefont {Redlich}},\ }\href
  {\doibase 10.1103/PhysRevC.83.054904} {\bibfield  {journal} {\bibinfo
  {journal} {Phys. Rev.}\ }\textbf {\bibinfo {volume} {C83}},\ \bibinfo {pages}
  {054904} (\bibinfo {year} {2011})},\ \Eprint {http://arxiv.org/abs/1008.4570}
  {arXiv:1008.4570 [hep-ph]} \BibitemShut {NoStop}%
%%CITATION = ARXIV:1008.4570;%%
\bibitem [{\citenamefont {Friman}\ \emph {et~al.}(2011)\citenamefont {Friman},
  \citenamefont {Karsch}, \citenamefont {Redlich},\ and\ \citenamefont
  {Skokov}}]{Friman:2011pf}%
  \BibitemOpen
  \bibfield  {author} {\bibinfo {author} {\bibfnamefont {B.}~\bibnamefont
  {Friman}}, \bibinfo {author} {\bibfnamefont {F.}~\bibnamefont {Karsch}},
  \bibinfo {author} {\bibfnamefont {K.}~\bibnamefont {Redlich}}, \ and\
  \bibinfo {author} {\bibfnamefont {V.}~\bibnamefont {Skokov}},\ }\href
  {\doibase 10.1140/epjc/s10052-011-1694-2} {\bibfield  {journal} {\bibinfo
  {journal} {Eur. Phys. J.}\ }\textbf {\bibinfo {volume} {C71}},\ \bibinfo
  {pages} {1694} (\bibinfo {year} {2011})},\ \Eprint
  {http://arxiv.org/abs/1103.3511} {arXiv:1103.3511 [hep-ph]} \BibitemShut
  {NoStop}%
%%CITATION = ARXIV:1103.3511;%%
\bibitem [{\citenamefont {Pisarski}\ and\ \citenamefont
  {Skokov}(2016)}]{Pisarski:2016ixt}%
  \BibitemOpen
  \bibfield  {author} {\bibinfo {author} {\bibfnamefont {R.~D.}\ \bibnamefont
  {Pisarski}}\ and\ \bibinfo {author} {\bibfnamefont {V.~V.}\ \bibnamefont
  {Skokov}},\ }\href {\doibase 10.1103/PhysRevD.94.034015} {\bibfield
  {journal} {\bibinfo  {journal} {Phys. Rev.}\ }\textbf {\bibinfo {volume}
  {D94}},\ \bibinfo {pages} {034015} (\bibinfo {year} {2016})},\ \Eprint
  {http://arxiv.org/abs/1604.00022} {arXiv:1604.00022 [hep-ph]} \BibitemShut
  {NoStop}%
%%CITATION = ARXIV:1604.00022;%%
\bibitem [{\citenamefont {Adamczyk}\ \emph {et~al.}(2014)\citenamefont
  {Adamczyk} \emph {et~al.}}]{Adamczyk:2013dal}%
  \BibitemOpen
  \bibfield  {author} {\bibinfo {author} {\bibfnamefont {L.}~\bibnamefont
  {Adamczyk}} \emph {et~al.} (\bibinfo {collaboration} {STAR}),\ }\href
  {\doibase 10.1103/PhysRevLett.112.032302} {\bibfield  {journal} {\bibinfo
  {journal} {Phys. Rev. Lett.}\ }\textbf {\bibinfo {volume} {112}},\ \bibinfo
  {pages} {032302} (\bibinfo {year} {2014})},\ \Eprint
  {http://arxiv.org/abs/1309.5681} {arXiv:1309.5681 [nucl-ex]} \BibitemShut
  {NoStop}%
%%CITATION = ARXIV:1309.5681;%%
\bibitem [{\citenamefont {Chattopadhyay}\ \emph {et~al.}(2016)\citenamefont
  {Chattopadhyay} \emph {et~al.}}]{Chattopadhyay:2016qhg}%
  \BibitemOpen
  \bibfield  {author} {\bibinfo {author} {\bibfnamefont {S.}~\bibnamefont
  {Chattopadhyay}} \emph {et~al.} (\bibinfo {collaboration} {CBM}),\
  }\href@noop {} {\  (\bibinfo {year} {2016})},\ \Eprint
  {http://arxiv.org/abs/1607.01487} {arXiv:1607.01487 [nucl-ex]} \BibitemShut
  {NoStop}%
%%CITATION = ARXIV:1607.01487;%%
\bibitem [{\citenamefont {Aoki}\ \emph {et~al.}(2006)\citenamefont {Aoki},
  \citenamefont {Endrodi}, \citenamefont {Fodor}, \citenamefont {Katz},\ and\
  \citenamefont {Szabo}}]{Aoki:2006we}%
  \BibitemOpen
  \bibfield  {author} {\bibinfo {author} {\bibfnamefont {Y.}~\bibnamefont
  {Aoki}}, \bibinfo {author} {\bibfnamefont {G.}~\bibnamefont {Endrodi}},
  \bibinfo {author} {\bibfnamefont {Z.}~\bibnamefont {Fodor}}, \bibinfo
  {author} {\bibfnamefont {S.~D.}\ \bibnamefont {Katz}}, \ and\ \bibinfo
  {author} {\bibfnamefont {K.~K.}\ \bibnamefont {Szabo}},\ }\href {\doibase
  10.1038/nature05120} {\bibfield  {journal} {\bibinfo  {journal} {Nature}\
  }\textbf {\bibinfo {volume} {443}},\ \bibinfo {pages} {675} (\bibinfo {year}
  {2006})},\ \Eprint {http://arxiv.org/abs/hep-lat/0611014}
  {arXiv:hep-lat/0611014 [hep-lat]} \BibitemShut {NoStop}%
%%CITATION = HEP-LAT/0611014;%%
\bibitem [{\citenamefont {Fodor}\ and\ \citenamefont
  {Katz}(2002)}]{Fodor:2001pe}%
  \BibitemOpen
  \bibfield  {author} {\bibinfo {author} {\bibfnamefont {Z.}~\bibnamefont
  {Fodor}}\ and\ \bibinfo {author} {\bibfnamefont {S.~D.}\ \bibnamefont
  {Katz}},\ }\href {\doibase 10.1088/1126-6708/2002/03/014} {\bibfield
  {journal} {\bibinfo  {journal} {JHEP}\ }\textbf {\bibinfo {volume} {03}},\
  \bibinfo {pages} {014} (\bibinfo {year} {2002})},\ \Eprint
  {http://arxiv.org/abs/hep-lat/0106002} {arXiv:hep-lat/0106002 [hep-lat]}
  \BibitemShut {NoStop}%
%%CITATION = HEP-LAT/0106002;%%
\bibitem [{\citenamefont {D'Elia}\ and\ \citenamefont
  {Lombardo}(2003)}]{D'Elia:2002gd}%
  \BibitemOpen
  \bibfield  {author} {\bibinfo {author} {\bibfnamefont {M.}~\bibnamefont
  {D'Elia}}\ and\ \bibinfo {author} {\bibfnamefont {M.-P.}\ \bibnamefont
  {Lombardo}},\ }\href {\doibase 10.1103/PhysRevD.67.014505} {\bibfield
  {journal} {\bibinfo  {journal} {Phys. Rev.}\ }\textbf {\bibinfo {volume}
  {D67}},\ \bibinfo {pages} {014505} (\bibinfo {year} {2003})},\ \Eprint
  {http://arxiv.org/abs/hep-lat/0209146} {arXiv:hep-lat/0209146 [hep-lat]}
  \BibitemShut {NoStop}%
%%CITATION = HEP-LAT/0209146;%%
\bibitem [{\citenamefont {Allton}\ \emph {et~al.}(2002)\citenamefont {Allton},
  \citenamefont {Ejiri}, \citenamefont {Hands}, \citenamefont {Kaczmarek},
  \citenamefont {Karsch}, \citenamefont {Laermann}, \citenamefont {Schmidt},\
  and\ \citenamefont {Scorzato}}]{Allton:2002zi}%
  \BibitemOpen
  \bibfield  {author} {\bibinfo {author} {\bibfnamefont {C.~R.}\ \bibnamefont
  {Allton}}, \bibinfo {author} {\bibfnamefont {S.}~\bibnamefont {Ejiri}},
  \bibinfo {author} {\bibfnamefont {S.~J.}\ \bibnamefont {Hands}}, \bibinfo
  {author} {\bibfnamefont {O.}~\bibnamefont {Kaczmarek}}, \bibinfo {author}
  {\bibfnamefont {F.}~\bibnamefont {Karsch}}, \bibinfo {author} {\bibfnamefont
  {E.}~\bibnamefont {Laermann}}, \bibinfo {author} {\bibfnamefont
  {C.}~\bibnamefont {Schmidt}}, \ and\ \bibinfo {author} {\bibfnamefont
  {L.}~\bibnamefont {Scorzato}},\ }\href {\doibase 10.1103/PhysRevD.66.074507}
  {\bibfield  {journal} {\bibinfo  {journal} {Phys. Rev.}\ }\textbf {\bibinfo
  {volume} {D66}},\ \bibinfo {pages} {074507} (\bibinfo {year} {2002})},\
  \Eprint {http://arxiv.org/abs/hep-lat/0204010} {arXiv:hep-lat/0204010
  [hep-lat]} \BibitemShut {NoStop}%
%%CITATION = HEP-LAT/0204010;%%
\bibitem [{\citenamefont {Fodor}\ and\ \citenamefont
  {Katz}(2004)}]{Fodor:2004nz}%
  \BibitemOpen
  \bibfield  {author} {\bibinfo {author} {\bibfnamefont {Z.}~\bibnamefont
  {Fodor}}\ and\ \bibinfo {author} {\bibfnamefont {S.~D.}\ \bibnamefont
  {Katz}},\ }\href {\doibase 10.1088/1126-6708/2004/04/050} {\bibfield
  {journal} {\bibinfo  {journal} {JHEP}\ }\textbf {\bibinfo {volume} {04}},\
  \bibinfo {pages} {050} (\bibinfo {year} {2004})},\ \Eprint
  {http://arxiv.org/abs/hep-lat/0402006} {arXiv:hep-lat/0402006 [hep-lat]}
  \BibitemShut {NoStop}%
%%CITATION = HEP-LAT/0402006;%%
\bibitem [{\citenamefont {Bonati}\ \emph {et~al.}(2014)\citenamefont {Bonati},
  \citenamefont {de~Forcrand}, \citenamefont {D'Elia}, \citenamefont
  {Philipsen},\ and\ \citenamefont {Sanfilippo}}]{Bonati:2014kpa}%
  \BibitemOpen
  \bibfield  {author} {\bibinfo {author} {\bibfnamefont {C.}~\bibnamefont
  {Bonati}}, \bibinfo {author} {\bibfnamefont {P.}~\bibnamefont {de~Forcrand}},
  \bibinfo {author} {\bibfnamefont {M.}~\bibnamefont {D'Elia}}, \bibinfo
  {author} {\bibfnamefont {O.}~\bibnamefont {Philipsen}}, \ and\ \bibinfo
  {author} {\bibfnamefont {F.}~\bibnamefont {Sanfilippo}},\ }\href {\doibase
  10.1103/PhysRevD.90.074030} {\bibfield  {journal} {\bibinfo  {journal} {Phys.
  Rev.}\ }\textbf {\bibinfo {volume} {D90}},\ \bibinfo {pages} {074030}
  (\bibinfo {year} {2014})},\ \Eprint {http://arxiv.org/abs/1408.5086}
  {arXiv:1408.5086 [hep-lat]} \BibitemShut {NoStop}%
%%CITATION = ARXIV:1408.5086;%%
\bibitem [{\citenamefont {de~Forcrand}\ and\ \citenamefont
  {Philipsen}(2010)}]{deForcrand:2010he}%
  \BibitemOpen
  \bibfield  {author} {\bibinfo {author} {\bibfnamefont {P.}~\bibnamefont
  {de~Forcrand}}\ and\ \bibinfo {author} {\bibfnamefont {O.}~\bibnamefont
  {Philipsen}},\ }\href {\doibase 10.1103/PhysRevLett.105.152001} {\bibfield
  {journal} {\bibinfo  {journal} {Phys. Rev. Lett.}\ }\textbf {\bibinfo
  {volume} {105}},\ \bibinfo {pages} {152001} (\bibinfo {year} {2010})},\
  \Eprint {http://arxiv.org/abs/1004.3144} {arXiv:1004.3144 [hep-lat]}
  \BibitemShut {NoStop}%
%%CITATION = ARXIV:1004.3144;%%
\bibitem [{\citenamefont {Stephanov}\ \emph {et~al.}(1999)\citenamefont
  {Stephanov}, \citenamefont {Rajagopal},\ and\ \citenamefont
  {Shuryak}}]{Stephanov:1999zu}%
  \BibitemOpen
  \bibfield  {author} {\bibinfo {author} {\bibfnamefont {M.~A.}\ \bibnamefont
  {Stephanov}}, \bibinfo {author} {\bibfnamefont {K.}~\bibnamefont
  {Rajagopal}}, \ and\ \bibinfo {author} {\bibfnamefont {E.~V.}\ \bibnamefont
  {Shuryak}},\ }\href {\doibase 10.1103/PhysRevD.60.114028} {\bibfield
  {journal} {\bibinfo  {journal} {Phys. Rev.}\ }\textbf {\bibinfo {volume}
  {D60}},\ \bibinfo {pages} {114028} (\bibinfo {year} {1999})},\ \Eprint
  {http://arxiv.org/abs/hep-ph/9903292} {arXiv:hep-ph/9903292 [hep-ph]}
  \BibitemShut {NoStop}%
%%CITATION = HEP-PH/9903292;%%
\bibitem [{\citenamefont {Ejiri}\ \emph {et~al.}(2006)\citenamefont {Ejiri},
  \citenamefont {Karsch},\ and\ \citenamefont {Redlich}}]{Ejiri:2005wq}%
  \BibitemOpen
  \bibfield  {author} {\bibinfo {author} {\bibfnamefont {S.}~\bibnamefont
  {Ejiri}}, \bibinfo {author} {\bibfnamefont {F.}~\bibnamefont {Karsch}}, \
  and\ \bibinfo {author} {\bibfnamefont {K.}~\bibnamefont {Redlich}},\ }\href
  {\doibase 10.1016/j.physletb.2005.11.083} {\bibfield  {journal} {\bibinfo
  {journal} {Phys. Lett.}\ }\textbf {\bibinfo {volume} {B633}},\ \bibinfo
  {pages} {275} (\bibinfo {year} {2006})},\ \Eprint
  {http://arxiv.org/abs/hep-ph/0509051} {arXiv:hep-ph/0509051 [hep-ph]}
  \BibitemShut {NoStop}%
%%CITATION = HEP-PH/0509051;%%
\bibitem [{\citenamefont {Stephanov}(2009)}]{Stephanov:2008qz}%
  \BibitemOpen
  \bibfield  {author} {\bibinfo {author} {\bibfnamefont {M.~A.}\ \bibnamefont
  {Stephanov}},\ }\href {\doibase 10.1103/PhysRevLett.102.032301} {\bibfield
  {journal} {\bibinfo  {journal} {Phys. Rev. Lett.}\ }\textbf {\bibinfo
  {volume} {102}},\ \bibinfo {pages} {032301} (\bibinfo {year} {2009})},\
  \Eprint {http://arxiv.org/abs/0809.3450} {arXiv:0809.3450 [hep-ph]}
  \BibitemShut {NoStop}%
%%CITATION = ARXIV:0809.3450;%%
\bibitem [{\citenamefont {Stephanov}(2011)}]{Stephanov:2011pb}%
  \BibitemOpen
  \bibfield  {author} {\bibinfo {author} {\bibfnamefont {M.~A.}\ \bibnamefont
  {Stephanov}},\ }\href {\doibase 10.1103/PhysRevLett.107.052301} {\bibfield
  {journal} {\bibinfo  {journal} {Phys. Rev. Lett.}\ }\textbf {\bibinfo
  {volume} {107}},\ \bibinfo {pages} {052301} (\bibinfo {year} {2011})},\
  \Eprint {http://arxiv.org/abs/1104.1627} {arXiv:1104.1627 [hep-ph]}
  \BibitemShut {NoStop}%
%%CITATION = ARXIV:1104.1627;%%
\bibitem [{\citenamefont {Bzdak}\ \emph {et~al.}(2013)\citenamefont {Bzdak},
  \citenamefont {Koch},\ and\ \citenamefont {Skokov}}]{Bzdak:2012an}%
  \BibitemOpen
  \bibfield  {author} {\bibinfo {author} {\bibfnamefont {A.}~\bibnamefont
  {Bzdak}}, \bibinfo {author} {\bibfnamefont {V.}~\bibnamefont {Koch}}, \ and\
  \bibinfo {author} {\bibfnamefont {V.}~\bibnamefont {Skokov}},\ }\href
  {\doibase 10.1103/PhysRevC.87.014901} {\bibfield  {journal} {\bibinfo
  {journal} {Phys. Rev.}\ }\textbf {\bibinfo {volume} {C87}},\ \bibinfo {pages}
  {014901} (\bibinfo {year} {2013})},\ \Eprint {http://arxiv.org/abs/1203.4529}
  {arXiv:1203.4529 [hep-ph]} \BibitemShut {NoStop}%
%%CITATION = ARXIV:1203.4529;%%
\bibitem [{\citenamefont {Skokov}\ \emph {et~al.}(2013)\citenamefont {Skokov},
  \citenamefont {Friman},\ and\ \citenamefont {Redlich}}]{Skokov:2012ds}%
  \BibitemOpen
  \bibfield  {author} {\bibinfo {author} {\bibfnamefont {V.}~\bibnamefont
  {Skokov}}, \bibinfo {author} {\bibfnamefont {B.}~\bibnamefont {Friman}}, \
  and\ \bibinfo {author} {\bibfnamefont {K.}~\bibnamefont {Redlich}},\ }\href
  {\doibase 10.1103/PhysRevC.88.034911} {\bibfield  {journal} {\bibinfo
  {journal} {Phys. Rev.}\ }\textbf {\bibinfo {volume} {C88}},\ \bibinfo {pages}
  {034911} (\bibinfo {year} {2013})},\ \Eprint {http://arxiv.org/abs/1205.4756}
  {arXiv:1205.4756 [hep-ph]} \BibitemShut {NoStop}%
%%CITATION = ARXIV:1205.4756;%%
\bibitem [{\citenamefont {Xu}(2016)}]{Xu:2016qzd}%
  \BibitemOpen
  \bibfield  {author} {\bibinfo {author} {\bibfnamefont {H.-j.}\ \bibnamefont
  {Xu}},\ }\href {\doibase 10.1103/PhysRevC.94.054903} {\bibfield  {journal}
  {\bibinfo  {journal} {Phys. Rev.}\ }\textbf {\bibinfo {volume} {C94}},\
  \bibinfo {pages} {054903} (\bibinfo {year} {2016})},\ \Eprint
  {http://arxiv.org/abs/1602.07089} {arXiv:1602.07089 [nucl-th]} \BibitemShut
  {NoStop}%
%%CITATION = ARXIV:1602.07089;%%
\bibitem [{\citenamefont {Braun-Munzinger}\ \emph {et~al.}(2016)\citenamefont
  {Braun-Munzinger}, \citenamefont {Rustamov},\ and\ \citenamefont
  {Stachel}}]{Braun-Munzinger:2016yjz}%
  \BibitemOpen
  \bibfield  {author} {\bibinfo {author} {\bibfnamefont {P.}~\bibnamefont
  {Braun-Munzinger}}, \bibinfo {author} {\bibfnamefont {A.}~\bibnamefont
  {Rustamov}}, \ and\ \bibinfo {author} {\bibfnamefont {J.}~\bibnamefont
  {Stachel}},\ }\href@noop {} {\  (\bibinfo {year} {2016})},\ \Eprint
  {http://arxiv.org/abs/1612.00702} {arXiv:1612.00702 [nucl-th]} \BibitemShut
  {NoStop}%
%%CITATION = ARXIV:1612.00702;%%
\bibitem [{\citenamefont {Bzdak}\ and\ \citenamefont
  {Koch}(2015)}]{Bzdak:2013pha}%
  \BibitemOpen
  \bibfield  {author} {\bibinfo {author} {\bibfnamefont {A.}~\bibnamefont
  {Bzdak}}\ and\ \bibinfo {author} {\bibfnamefont {V.}~\bibnamefont {Koch}},\
  }\href {\doibase 10.1103/PhysRevC.91.027901} {\bibfield  {journal} {\bibinfo
  {journal} {Phys. Rev.}\ }\textbf {\bibinfo {volume} {C91}},\ \bibinfo {pages}
  {027901} (\bibinfo {year} {2015})},\ \Eprint {http://arxiv.org/abs/1312.4574}
  {arXiv:1312.4574 [nucl-th]} \BibitemShut {NoStop}%
%%CITATION = ARXIV:1312.4574;%%
\bibitem [{\citenamefont {Ling}\ and\ \citenamefont
  {Stephanov}(2016)}]{Ling:2015yau}%
  \BibitemOpen
  \bibfield  {author} {\bibinfo {author} {\bibfnamefont {B.}~\bibnamefont
  {Ling}}\ and\ \bibinfo {author} {\bibfnamefont {M.~A.}\ \bibnamefont
  {Stephanov}},\ }\href {\doibase 10.1103/PhysRevC.93.034915} {\bibfield
  {journal} {\bibinfo  {journal} {Phys. Rev.}\ }\textbf {\bibinfo {volume}
  {C93}},\ \bibinfo {pages} {034915} (\bibinfo {year} {2016})},\ \Eprint
  {http://arxiv.org/abs/1512.09125} {arXiv:1512.09125 [nucl-th]} \BibitemShut
  {NoStop}%
%%CITATION = ARXIV:1512.09125;%%
\bibitem [{\citenamefont {Bzdak}\ \emph
  {et~al.}(2016{\natexlab{a}})\citenamefont {Bzdak}, \citenamefont {Holzmann},\
  and\ \citenamefont {Koch}}]{Bzdak:2016qdc}%
  \BibitemOpen
  \bibfield  {author} {\bibinfo {author} {\bibfnamefont {A.}~\bibnamefont
  {Bzdak}}, \bibinfo {author} {\bibfnamefont {R.}~\bibnamefont {Holzmann}}, \
  and\ \bibinfo {author} {\bibfnamefont {V.}~\bibnamefont {Koch}},\ }\href@noop
  {} {\  (\bibinfo {year} {2016}{\natexlab{a}})},\ \Eprint
  {http://arxiv.org/abs/1603.09057} {arXiv:1603.09057 [nucl-th]} \BibitemShut
  {NoStop}%
%%CITATION = ARXIV:1603.09057;%%
\bibitem [{\citenamefont {Luo}(2015{\natexlab{a}})}]{Luo:2014rea}%
  \BibitemOpen
  \bibfield  {author} {\bibinfo {author} {\bibfnamefont {X.}~\bibnamefont
  {Luo}},\ }\href {\doibase 10.1103/PhysRevC.94.059901,
  10.1103/PhysRevC.91.034907} {\bibfield  {journal} {\bibinfo  {journal} {Phys.
  Rev.}\ }\textbf {\bibinfo {volume} {C91}},\ \bibinfo {pages} {034907}
  (\bibinfo {year} {2015}{\natexlab{a}})},\ \bibinfo {note} {[Erratum: Phys.
  Rev.C94,no.5,059901(2016)]},\ \Eprint {http://arxiv.org/abs/1410.3914}
  {arXiv:1410.3914 [physics.data-an]} \BibitemShut {NoStop}%
%%CITATION = ARXIV:1410.3914;%%
\bibitem [{\citenamefont {Kitazawa}(2016)}]{Kitazawa:2016awu}%
  \BibitemOpen
  \bibfield  {author} {\bibinfo {author} {\bibfnamefont {M.}~\bibnamefont
  {Kitazawa}},\ }\href {\doibase 10.1103/PhysRevC.93.044911} {\bibfield
  {journal} {\bibinfo  {journal} {Phys. Rev.}\ }\textbf {\bibinfo {volume}
  {C93}},\ \bibinfo {pages} {044911} (\bibinfo {year} {2016})},\ \Eprint
  {http://arxiv.org/abs/1602.01234} {arXiv:1602.01234 [nucl-th]} \BibitemShut
  {NoStop}%
%%CITATION = ARXIV:1602.01234;%%
\bibitem [{\citenamefont {Nonaka}\ \emph {et~al.}(2016)\citenamefont {Nonaka},
  \citenamefont {Sugiura}, \citenamefont {Esumi}, \citenamefont {Masui},\ and\
  \citenamefont {Luo}}]{Nonaka:2016xje}%
  \BibitemOpen
  \bibfield  {author} {\bibinfo {author} {\bibfnamefont {T.}~\bibnamefont
  {Nonaka}}, \bibinfo {author} {\bibfnamefont {T.}~\bibnamefont {Sugiura}},
  \bibinfo {author} {\bibfnamefont {S.}~\bibnamefont {Esumi}}, \bibinfo
  {author} {\bibfnamefont {H.}~\bibnamefont {Masui}}, \ and\ \bibinfo {author}
  {\bibfnamefont {X.}~\bibnamefont {Luo}},\ }\href {\doibase
  10.1103/PhysRevC.94.034909} {\bibfield  {journal} {\bibinfo  {journal} {Phys.
  Rev.}\ }\textbf {\bibinfo {volume} {C94}},\ \bibinfo {pages} {034909}
  (\bibinfo {year} {2016})},\ \Eprint {http://arxiv.org/abs/1604.06212}
  {arXiv:1604.06212 [nucl-th]} \BibitemShut {NoStop}%
%%CITATION = ARXIV:1604.06212;%%
\bibitem [{\citenamefont {Kitazawa}\ and\ \citenamefont
  {Asakawa}(2012{\natexlab{a}})}]{Kitazawa:2011wh}%
  \BibitemOpen
  \bibfield  {author} {\bibinfo {author} {\bibfnamefont {M.}~\bibnamefont
  {Kitazawa}}\ and\ \bibinfo {author} {\bibfnamefont {M.}~\bibnamefont
  {Asakawa}},\ }\href {\doibase 10.1103/PhysRevC.85.021901} {\bibfield
  {journal} {\bibinfo  {journal} {Phys. Rev.}\ }\textbf {\bibinfo {volume}
  {C85}},\ \bibinfo {pages} {021901} (\bibinfo {year} {2012}{\natexlab{a}})},\
  \Eprint {http://arxiv.org/abs/1107.2755} {arXiv:1107.2755 [nucl-th]}
  \BibitemShut {NoStop}%
%%CITATION = ARXIV:1107.2755;%%
\bibitem [{\citenamefont {Fecková}\ \emph {et~al.}(2015)\citenamefont
  {Fecková}, \citenamefont {Steinheimer}, \citenamefont {Tomášik},\ and\
  \citenamefont {Bleicher}}]{Feckova:2015qza}%
  \BibitemOpen
  \bibfield  {author} {\bibinfo {author} {\bibfnamefont {Z.}~\bibnamefont
  {Fecková}}, \bibinfo {author} {\bibfnamefont {J.}~\bibnamefont
  {Steinheimer}}, \bibinfo {author} {\bibfnamefont {B.}~\bibnamefont
  {Tomášik}}, \ and\ \bibinfo {author} {\bibfnamefont {M.}~\bibnamefont
  {Bleicher}},\ }\href {\doibase 10.1103/PhysRevC.92.064908} {\bibfield
  {journal} {\bibinfo  {journal} {Phys. Rev.}\ }\textbf {\bibinfo {volume}
  {C92}},\ \bibinfo {pages} {064908} (\bibinfo {year} {2015})},\ \Eprint
  {http://arxiv.org/abs/1510.05519} {arXiv:1510.05519 [nucl-th]} \BibitemShut
  {NoStop}%
%%CITATION = ARXIV:1510.05519;%%
\bibitem [{\citenamefont {Mukherjee}\ \emph {et~al.}(2015)\citenamefont
  {Mukherjee}, \citenamefont {Venugopalan},\ and\ \citenamefont
  {Yin}}]{Mukherjee:2015swa}%
  \BibitemOpen
  \bibfield  {author} {\bibinfo {author} {\bibfnamefont {S.}~\bibnamefont
  {Mukherjee}}, \bibinfo {author} {\bibfnamefont {R.}~\bibnamefont
  {Venugopalan}}, \ and\ \bibinfo {author} {\bibfnamefont {Y.}~\bibnamefont
  {Yin}},\ }\href {\doibase 10.1103/PhysRevC.92.034912} {\bibfield  {journal}
  {\bibinfo  {journal} {Phys. Rev.}\ }\textbf {\bibinfo {volume} {C92}},\
  \bibinfo {pages} {034912} (\bibinfo {year} {2015})},\ \Eprint
  {http://arxiv.org/abs/1506.00645} {arXiv:1506.00645 [hep-ph]} \BibitemShut
  {NoStop}%
%%CITATION = ARXIV:1506.00645;%%
\bibitem [{\citenamefont {Mukherjee}\ \emph {et~al.}(2016)\citenamefont
  {Mukherjee}, \citenamefont {Venugopalan},\ and\ \citenamefont
  {Yin}}]{Mukherjee:2016kyu}%
  \BibitemOpen
  \bibfield  {author} {\bibinfo {author} {\bibfnamefont {S.}~\bibnamefont
  {Mukherjee}}, \bibinfo {author} {\bibfnamefont {R.}~\bibnamefont
  {Venugopalan}}, \ and\ \bibinfo {author} {\bibfnamefont {Y.}~\bibnamefont
  {Yin}},\ }\href {\doibase 10.1103/PhysRevLett.117.222301} {\bibfield
  {journal} {\bibinfo  {journal} {Phys. Rev. Lett.}\ }\textbf {\bibinfo
  {volume} {117}},\ \bibinfo {pages} {222301} (\bibinfo {year} {2016})},\
  \Eprint {http://arxiv.org/abs/1605.09341} {arXiv:1605.09341 [hep-ph]}
  \BibitemShut {NoStop}%
%%CITATION = ARXIV:1605.09341;%%
\bibitem [{\citenamefont {Bialas}\ \emph {et~al.}(2016)\citenamefont {Bialas},
  \citenamefont {Bzdak},\ and\ \citenamefont {Koch}}]{Bialas:2016epd}%
  \BibitemOpen
  \bibfield  {author} {\bibinfo {author} {\bibfnamefont {A.}~\bibnamefont
  {Bialas}}, \bibinfo {author} {\bibfnamefont {A.}~\bibnamefont {Bzdak}}, \
  and\ \bibinfo {author} {\bibfnamefont {V.}~\bibnamefont {Koch}},\ }\href@noop
  {} {\  (\bibinfo {year} {2016})},\ \Eprint {http://arxiv.org/abs/1608.07041}
  {arXiv:1608.07041 [hep-ph]} \BibitemShut {NoStop}%
%%CITATION = ARXIV:1608.07041;%%
\bibitem [{\citenamefont {Bzdak}\ \emph
  {et~al.}(2016{\natexlab{b}})\citenamefont {Bzdak}, \citenamefont {Koch},\
  and\ \citenamefont {Strodthoff}}]{Bzdak:2016sxg}%
  \BibitemOpen
  \bibfield  {author} {\bibinfo {author} {\bibfnamefont {A.}~\bibnamefont
  {Bzdak}}, \bibinfo {author} {\bibfnamefont {V.}~\bibnamefont {Koch}}, \ and\
  \bibinfo {author} {\bibfnamefont {N.}~\bibnamefont {Strodthoff}},\
  }\href@noop {} {\  (\bibinfo {year} {2016}{\natexlab{b}})},\ \Eprint
  {http://arxiv.org/abs/1607.07375} {arXiv:1607.07375 [nucl-th]} \BibitemShut
  {NoStop}%
%%CITATION = ARXIV:1607.07375;%%
\bibitem [{\citenamefont {Bialas}\ \emph {et~al.}(1976)\citenamefont {Bialas},
  \citenamefont {Bleszynski},\ and\ \citenamefont {Czyz}}]{Bialas:1976ed}%
  \BibitemOpen
  \bibfield  {author} {\bibinfo {author} {\bibfnamefont {A.}~\bibnamefont
  {Bialas}}, \bibinfo {author} {\bibfnamefont {M.}~\bibnamefont {Bleszynski}},
  \ and\ \bibinfo {author} {\bibfnamefont {W.}~\bibnamefont {Czyz}},\ }\href
  {\doibase 10.1016/0550-3213(76)90329-1} {\bibfield  {journal} {\bibinfo
  {journal} {Nucl. Phys.}\ }\textbf {\bibinfo {volume} {B111}},\ \bibinfo
  {pages} {461} (\bibinfo {year} {1976})}\BibitemShut {NoStop}%
%%CITATION = NUPHA,B111,461;%%
\bibitem [{\citenamefont {Luo}(2015{\natexlab{b}})}]{Luo:2015ewa}%
  \BibitemOpen
  \bibfield  {author} {\bibinfo {author} {\bibfnamefont {X.}~\bibnamefont
  {Luo}} (\bibinfo {collaboration} {STAR}),\ }\bibfield  {booktitle} {\emph
  {\bibinfo {booktitle} {{Proceedings, 9th International Workshop on Critical
  Point and Onset of Deconfinement (CPOD 2014): Bielefeld, Germany, November
  17-21, 2014}}},\ }\href@noop {} {\bibfield  {journal} {\bibinfo  {journal}
  {PoS}\ }\textbf {\bibinfo {volume} {CPOD2014}},\ \bibinfo {pages} {019}
  (\bibinfo {year} {2015}{\natexlab{b}})},\ \Eprint
  {http://arxiv.org/abs/1503.02558} {arXiv:1503.02558 [nucl-ex]} \BibitemShut
  {NoStop}%
%%CITATION = ARXIV:1503.02558;%%
\bibitem [{\citenamefont {Luo}(2016)}]{Luo:2015doi}%
  \BibitemOpen
  \bibfield  {author} {\bibinfo {author} {\bibfnamefont {X.}~\bibnamefont
  {Luo}},\ }\bibfield  {booktitle} {\emph {\bibinfo {booktitle} {{Proceedings,
  25th International Conference on Ultra-Relativistic Nucleus-Nucleus
  Collisions (Quark Matter 2015): Kobe, Japan, September 27-October 3,
  2015}}},\ }\href {\doibase 10.1016/j.nuclphysa.2016.03.025} {\bibfield
  {journal} {\bibinfo  {journal} {Nucl. Phys.}\ }\textbf {\bibinfo {volume}
  {A956}},\ \bibinfo {pages} {75} (\bibinfo {year} {2016})},\ \Eprint
  {http://arxiv.org/abs/1512.09215} {arXiv:1512.09215 [nucl-ex]} \BibitemShut
  {NoStop}%
%%CITATION = ARXIV:1512.09215;%%
\bibitem [{\citenamefont {Koch}\ \emph {et~al.}(2001)\citenamefont {Koch},
  \citenamefont {Bleicher},\ and\ \citenamefont {Jeon}}]{Koch:2001cb}%
  \BibitemOpen
  \bibfield  {author} {\bibinfo {author} {\bibfnamefont {V.}~\bibnamefont
  {Koch}}, \bibinfo {author} {\bibfnamefont {M.}~\bibnamefont {Bleicher}}, \
  and\ \bibinfo {author} {\bibfnamefont {S.}~\bibnamefont {Jeon}},\ }\bibfield
  {booktitle} {\emph {\bibinfo {booktitle} {{Fundamental issues in elementary
  matter. Proceedings, Symposium, 241st WE-Heraeus Seminar, Bad Honnef,
  Germany, September 25-29, 2000}}},\ }\href {\doibase
  10.1556/APH.14.2001.1-4.22} {\bibfield  {journal} {\bibinfo  {journal} {Heavy
  Ion Phys.}\ }\textbf {\bibinfo {volume} {14}},\ \bibinfo {pages} {227}
  (\bibinfo {year} {2001})}\BibitemShut {NoStop}%
%%CITATION = APHPF,14,227;%%
\bibitem [{\citenamefont {Koch}(2010)}]{Koch:2008ia}%
  \BibitemOpen
  \bibfield  {author} {\bibinfo {author} {\bibfnamefont {V.}~\bibnamefont
  {Koch}},\ }in\ \href {\doibase 10.1007/978-3-642-01539-7_20} {\emph {\bibinfo
  {booktitle} {Relativistic Heavy Ion Physics}}},\ \bibinfo {series}
  {Landolt-Boernstein New Series I}, Vol.~\bibinfo {volume} {23},\ \bibinfo
  {editor} {edited by\ \bibinfo {editor} {\bibfnamefont {R.}~\bibnamefont
  {Stock}}}\ (\bibinfo  {publisher} {Springer},\ \bibinfo {address}
  {Heidelberg},\ \bibinfo {year} {2010})\ pp.\ \bibinfo {pages} {626--652},\
  \Eprint {http://arxiv.org/abs/0810.2520} {arXiv:0810.2520 [nucl-th]}
  \BibitemShut {NoStop}%
%%CITATION = ARXIV:0810.2520;%%
\bibitem [{\citenamefont {Kitazawa}\ and\ \citenamefont
  {Asakawa}(2012{\natexlab{b}})}]{Kitazawa:2012at}%
  \BibitemOpen
  \bibfield  {author} {\bibinfo {author} {\bibfnamefont {M.}~\bibnamefont
  {Kitazawa}}\ and\ \bibinfo {author} {\bibfnamefont {M.}~\bibnamefont
  {Asakawa}},\ }\href {\doibase 10.1103/PhysRevC.86.024904,
  10.1103/PhysRevC.86.069902} {\bibfield  {journal} {\bibinfo  {journal} {Phys.
  Rev.}\ }\textbf {\bibinfo {volume} {C86}},\ \bibinfo {pages} {024904}
  (\bibinfo {year} {2012}{\natexlab{b}})},\ \bibinfo {note} {[Erratum: Phys.
  Rev.C86,069902(2012)]},\ \Eprint {http://arxiv.org/abs/1205.3292}
  {arXiv:1205.3292 [nucl-th]} \BibitemShut {NoStop}%
%%CITATION = ARXIV:1205.3292;%%
\bibitem [{\citenamefont {Alver}\ \emph {et~al.}(2008)\citenamefont {Alver},
  \citenamefont {Baker}, \citenamefont {Loizides},\ and\ \citenamefont
  {Steinberg}}]{Alver:2008aq}%
  \BibitemOpen
  \bibfield  {author} {\bibinfo {author} {\bibfnamefont {B.}~\bibnamefont
  {Alver}}, \bibinfo {author} {\bibfnamefont {M.}~\bibnamefont {Baker}},
  \bibinfo {author} {\bibfnamefont {C.}~\bibnamefont {Loizides}}, \ and\
  \bibinfo {author} {\bibfnamefont {P.}~\bibnamefont {Steinberg}},\ }\href@noop
  {} {\  (\bibinfo {year} {2008})},\ \Eprint {http://arxiv.org/abs/0805.4411}
  {arXiv:0805.4411 [nucl-ex]} \BibitemShut {NoStop}%
%%CITATION = ARXIV:0805.4411;%%
\bibitem [{\citenamefont {Basile}\ \emph {et~al.}(1983)\citenamefont {Basile}
  \emph {et~al.}}]{Basile:1982we}%
  \BibitemOpen
  \bibfield  {author} {\bibinfo {author} {\bibfnamefont {M.}~\bibnamefont
  {Basile}} \emph {et~al.},\ }\bibfield  {booktitle} {\emph {\bibinfo
  {booktitle} {{Ferrara International School Niccolò Cabeo 2014 Ferrara,
  Italy, May 19-23, 2014}}},\ }\href {\doibase 10.1007/BF02724233} {\bibfield
  {journal} {\bibinfo  {journal} {Nuovo Cim.}\ }\textbf {\bibinfo {volume}
  {A73}},\ \bibinfo {pages} {329} (\bibinfo {year} {1983})}\BibitemShut
  {NoStop}%
%%CITATION = NUCIA,A73,329;%%
\bibitem [{\citenamefont {Pruneau}\ \emph {et~al.}(2002)\citenamefont
  {Pruneau}, \citenamefont {Gavin},\ and\ \citenamefont
  {Voloshin}}]{Pruneau:2002yf}%
  \BibitemOpen
  \bibfield  {author} {\bibinfo {author} {\bibfnamefont {C.}~\bibnamefont
  {Pruneau}}, \bibinfo {author} {\bibfnamefont {S.}~\bibnamefont {Gavin}}, \
  and\ \bibinfo {author} {\bibfnamefont {S.}~\bibnamefont {Voloshin}},\ }\href
  {\doibase 10.1103/PhysRevC.66.044904} {\bibfield  {journal} {\bibinfo
  {journal} {Phys. Rev.}\ }\textbf {\bibinfo {volume} {C66}},\ \bibinfo {pages}
  {044904} (\bibinfo {year} {2002})},\ \Eprint
  {http://arxiv.org/abs/nucl-ex/0204011} {arXiv:nucl-ex/0204011 [nucl-ex]}
  \BibitemShut {NoStop}%
%%CITATION = NUCL-EX/0204011;%%
\bibitem [{\citenamefont {Grosse-Oetringhaus}\ and\ \citenamefont
  {Reygers}(2010)}]{GrosseOetringhaus:2009kz}%
  \BibitemOpen
  \bibfield  {author} {\bibinfo {author} {\bibfnamefont {J.~F.}\ \bibnamefont
  {Grosse-Oetringhaus}}\ and\ \bibinfo {author} {\bibfnamefont
  {K.}~\bibnamefont {Reygers}},\ }\href {\doibase
  10.1088/0954-3899/37/8/083001} {\bibfield  {journal} {\bibinfo  {journal} {J.
  Phys.}\ }\textbf {\bibinfo {volume} {G37}},\ \bibinfo {pages} {083001}
  (\bibinfo {year} {2010})},\ \Eprint {http://arxiv.org/abs/0912.0023}
  {arXiv:0912.0023 [hep-ex]} \BibitemShut {NoStop}%
%%CITATION = ARXIV:0912.0023;%%
\bibitem [{\citenamefont {Adcox}\ \emph {et~al.}(2001)\citenamefont {Adcox}
  \emph {et~al.}}]{Adcox:2000sp}%
  \BibitemOpen
  \bibfield  {author} {\bibinfo {author} {\bibfnamefont {K.}~\bibnamefont
  {Adcox}} \emph {et~al.} (\bibinfo {collaboration} {PHENIX}),\ }\href
  {\doibase 10.1103/PhysRevLett.86.3500} {\bibfield  {journal} {\bibinfo
  {journal} {Phys. Rev. Lett.}\ }\textbf {\bibinfo {volume} {86}},\ \bibinfo
  {pages} {3500} (\bibinfo {year} {2001})},\ \Eprint
  {http://arxiv.org/abs/nucl-ex/0012008} {arXiv:nucl-ex/0012008 [nucl-ex]}
  \BibitemShut {NoStop}%
%%CITATION = NUCL-EX/0012008;%%
\bibitem [{\citenamefont {Back}\ \emph {et~al.}(2006)\citenamefont {Back} \emph
  {et~al.}}]{Back:2005hs}%
  \BibitemOpen
  \bibfield  {author} {\bibinfo {author} {\bibfnamefont {B.~B.}\ \bibnamefont
  {Back}} \emph {et~al.} (\bibinfo {collaboration} {PHOBOS}),\ }\href {\doibase
  10.1103/PhysRevC.74.021901} {\bibfield  {journal} {\bibinfo  {journal} {Phys.
  Rev.}\ }\textbf {\bibinfo {volume} {C74}},\ \bibinfo {pages} {021901}
  (\bibinfo {year} {2006})},\ \Eprint {http://arxiv.org/abs/nucl-ex/0509034}
  {arXiv:nucl-ex/0509034 [nucl-ex]} \BibitemShut {NoStop}%
%%CITATION = NUCL-EX/0509034;%%
\bibitem [{\citenamefont {Eremin}\ and\ \citenamefont
  {Voloshin}(2003)}]{Eremin:2003qn}%
  \BibitemOpen
  \bibfield  {author} {\bibinfo {author} {\bibfnamefont {S.}~\bibnamefont
  {Eremin}}\ and\ \bibinfo {author} {\bibfnamefont {S.}~\bibnamefont
  {Voloshin}},\ }\href {\doibase 10.1103/PhysRevC.67.064905} {\bibfield
  {journal} {\bibinfo  {journal} {Phys. Rev.}\ }\textbf {\bibinfo {volume}
  {C67}},\ \bibinfo {pages} {064905} (\bibinfo {year} {2003})},\ \Eprint
  {http://arxiv.org/abs/nucl-th/0302071} {arXiv:nucl-th/0302071 [nucl-th]}
  \BibitemShut {NoStop}%
%%CITATION = NUCL-TH/0302071;%%
\bibitem [{\citenamefont {Bialas}\ and\ \citenamefont
  {Bzdak}(2007)}]{Bialas:2006kw}%
  \BibitemOpen
  \bibfield  {author} {\bibinfo {author} {\bibfnamefont {A.}~\bibnamefont
  {Bialas}}\ and\ \bibinfo {author} {\bibfnamefont {A.}~\bibnamefont {Bzdak}},\
  }\href {\doibase 10.1016/j.physletb.2007.04.014} {\bibfield  {journal}
  {\bibinfo  {journal} {Phys. Lett.}\ }\textbf {\bibinfo {volume} {B649}},\
  \bibinfo {pages} {263} (\bibinfo {year} {2007})},\ \Eprint
  {http://arxiv.org/abs/nucl-th/0611021} {arXiv:nucl-th/0611021 [nucl-th]}
  \BibitemShut {NoStop}%
%%CITATION = NUCL-TH/0611021;%%
\bibitem [{\citenamefont {Skokov}\ and\ \citenamefont
  {Voskresensky}(2009{\natexlab{a}})}]{Skokov:2008zp}%
  \BibitemOpen
  \bibfield  {author} {\bibinfo {author} {\bibfnamefont {V.~V.}\ \bibnamefont
  {Skokov}}\ and\ \bibinfo {author} {\bibfnamefont {D.~N.}\ \bibnamefont
  {Voskresensky}},\ }\href {\doibase 10.1134/S0021364009160012} {\bibfield
  {journal} {\bibinfo  {journal} {JETP Lett.}\ }\textbf {\bibinfo {volume}
  {90}},\ \bibinfo {pages} {223} (\bibinfo {year} {2009}{\natexlab{a}})},\
  \Eprint {http://arxiv.org/abs/0811.3868} {arXiv:0811.3868 [nucl-th]}
  \BibitemShut {NoStop}%
%%CITATION = ARXIV:0811.3868;%%
\bibitem [{\citenamefont {Skokov}\ and\ \citenamefont
  {Voskresensky}(2009{\natexlab{b}})}]{Skokov:2009yu}%
  \BibitemOpen
  \bibfield  {author} {\bibinfo {author} {\bibfnamefont {V.~V.}\ \bibnamefont
  {Skokov}}\ and\ \bibinfo {author} {\bibfnamefont {D.~N.}\ \bibnamefont
  {Voskresensky}},\ }\href {\doibase 10.1016/j.nuclphysa.2009.07.012}
  {\bibfield  {journal} {\bibinfo  {journal} {Nucl. Phys.}\ }\textbf {\bibinfo
  {volume} {A828}},\ \bibinfo {pages} {401} (\bibinfo {year}
  {2009}{\natexlab{b}})},\ \Eprint {http://arxiv.org/abs/0903.4335}
  {arXiv:0903.4335 [nucl-th]} \BibitemShut {NoStop}%
%%CITATION = ARXIV:0903.4335;%%
\bibitem [{\citenamefont {Steinheimer}\ and\ \citenamefont
  {Randrup}(2012)}]{Steinheimer:2012gc}%
  \BibitemOpen
  \bibfield  {author} {\bibinfo {author} {\bibfnamefont {J.}~\bibnamefont
  {Steinheimer}}\ and\ \bibinfo {author} {\bibfnamefont {J.}~\bibnamefont
  {Randrup}},\ }\href {\doibase 10.1103/PhysRevLett.109.212301} {\bibfield
  {journal} {\bibinfo  {journal} {Phys. Rev. Lett.}\ }\textbf {\bibinfo
  {volume} {109}},\ \bibinfo {pages} {212301} (\bibinfo {year} {2012})},\
  \Eprint {http://arxiv.org/abs/1209.2462} {arXiv:1209.2462 [nucl-th]}
  \BibitemShut {NoStop}%
%%CITATION = ARXIV:1209.2462;%%
\bibitem [{\citenamefont {Steinheimer}\ \emph {et~al.}(2014)\citenamefont
  {Steinheimer}, \citenamefont {Randrup},\ and\ \citenamefont
  {Koch}}]{Steinheimer:2013xxa}%
  \BibitemOpen
  \bibfield  {author} {\bibinfo {author} {\bibfnamefont {J.}~\bibnamefont
  {Steinheimer}}, \bibinfo {author} {\bibfnamefont {J.}~\bibnamefont
  {Randrup}}, \ and\ \bibinfo {author} {\bibfnamefont {V.}~\bibnamefont
  {Koch}},\ }\href {\doibase 10.1103/PhysRevC.89.034901} {\bibfield  {journal}
  {\bibinfo  {journal} {Phys. Rev.}\ }\textbf {\bibinfo {volume} {C89}},\
  \bibinfo {pages} {034901} (\bibinfo {year} {2014})},\ \Eprint
  {http://arxiv.org/abs/1311.0999} {arXiv:1311.0999 [nucl-th]} \BibitemShut
  {NoStop}%
%%CITATION = ARXIV:1311.0999;%%
\bibitem [{\citenamefont {Pumplin}(1994)}]{Pumplin:1994gh}%
  \BibitemOpen
  \bibfield  {author} {\bibinfo {author} {\bibfnamefont {J.}~\bibnamefont
  {Pumplin}},\ }\href {\doibase 10.1103/PhysRevD.50.6811} {\bibfield  {journal}
  {\bibinfo  {journal} {Phys. Rev.}\ }\textbf {\bibinfo {volume} {D50}},\
  \bibinfo {pages} {6811} (\bibinfo {year} {1994})},\ \Eprint
  {http://arxiv.org/abs/hep-ph/9407332} {arXiv:hep-ph/9407332 [hep-ph]}
  \BibitemShut {NoStop}%
%%CITATION = HEP-PH/9407332;%%
\end{thebibliography}%

%\begin{thebibliography}{99}

%\end{thebibliography}

\end{document}